\documentclass[useAMS,usenatbib,onecolumn]{mn2e}
\usepackage{graphicx}
\usepackage{bm}
\usepackage{ulem}

\newcommand{\bea}{\begin{eqnarray}}
\newcommand{\eea}{\end{eqnarray}}
\newcommand{\beq}{\begin{equation}}
\newcommand{\eeq}{\end{equation}}

\newcommand{\simless}[0]{\mathbin{\lower 3pt\hbox
   {$\rlap{\raise 5pt\hbox{$\char'074$}}\mathchar"7218$}}}
\newcommand{\simgreat}[0]{\mathbin{\lower 3pt\hbox
   {$\rlap{\raise 5pt\hbox{$\char'076$}}\mathchar"7218$}}}

\newcommand{\figref}[1]{figure \ref{#1}}

\newcommand{\capfigref}[1]{Figure \ref{#1}}

\newcommand{\eqnref}[1]{eq. (\ref{#1})}
\newcommand{\eqnrefs}[1]{eqs. (\ref{#1})}
\newcommand{\eqnrefbare}[1]{(\ref{#1})}
\newcommand{\capeqnref}[1]{Equation (\ref{#1})}
\newcommand{\capeqnrefs}[1]{Equations (\ref{#1})}
\graphicspath{{}{figures/}}
\title[Non-linear density-velocity relation]{Non-linear density-velocity divergence relation from phase space dynamics.}
\author[Sharvari Nadkarni-Ghosh]{Sharvari Nadkarni-Ghosh$^{1,}$$^{2}$\thanks{E-mail:
nsharvari@gmail.com, sharvari@iitk.ac.in}
\\ $^{1,}$Department of Theoretical Sciences, S. N. Bose National Centre for Basic Sciences, Sector-III, Block-JD, Salt Lake, Kolkata-700098 India
\\ $^{2}$Department of Physics, IIT Kanpur, Kanpur, U.P. 208016 India } 

\begin{document}
\date{}
\pagerange{\pageref{firstpage}--\pageref{lastpage}} \pubyear{}

\maketitle
\label{firstpage}

\begin{abstract}

We obtain the non-linear relation between cosmological density and velocity perturbations by examining their 
joint dynamics in a two dimensional density-velocity divergence phase space. We restrict to spatially flat cosmologies consisting of pressureless matter and non-clustering dark energy characterised by a constant equation of state $w$. Using the spherical top-hat model, we derive the coupled equations that govern the joint evolution of density and velocity perturbations and examine the flow generated by this system. In general, the initial density and velocity are independent, but requiring that the perturbations vanish at the big bang time sets a relation between the two. This traces out a curve in the instantaneous phase space, which we call the `Zel'dovich curve'. We show that this curve acts like an attracting solution for the phase space dynamics and is the desired non-linear extension of the density-velocity divergence relation. We obtain a fitting formula which is a combination of the formulae by Bernardeau and Bilicki \& Chodorowski, generalised to include the dependence on $w$. We find that as in the linear regime, the explicit dependence on the dark energy equation of state stays weak even in the non-linear regime.Although the result itself is somewhat expected, the new feature of this work is the interpretation of the relation in the phase space picture and the generality of the method. Finally,  as an observational implication, we examine the evolution of galaxy cluster profiles using the spherical infall model for different values of $w$. We demonstrate that using only the density or the velocity information to constrain $w$ is subject to degeneracies in other parameters such as $\sigma_8$ but plotting observations onto the joint density-velocity phase space can help lift this degeneracy.

\end{abstract}

\section{Introduction}

The distribution of matter on very large scales ($\sim 100$ Mpc) is fairly homogenous (e.g., \citealt{hogg_cosmic_2005}, \citealt{sarkar_scale_2009}, \citealt{scrimgeour_wigglez_2012}), but on smaller scales it is far from it. The fractional overdensity ($\delta$) and the peculiar velocity (${\bf v}$) are the two variables that characterise this inhomogeneity and are related via the continuity equation, the Euler equation and the law of gravitation. When the inhomogeneities are small, the equations can be linearised and ignoring the decaying mode results in the local relation $\nabla \cdot {\bf v} = -f H\delta$, where $H$ is the Hubble parameter and $f$ is the linear growth factor. In the theory of gravitational instability, except in the orbit-crossing regions, the peculiar velocity is curl free and the above equation completely characterises the relation between the two fields in the linear regime. $f$ mainly depends on the matter density parameter $\Omega_m$, and is usually expressed as $\Omega_m^\gamma$, where $\gamma$ is the growth index.  Given $\gamma$, a comparison of the data from redshift and peculiar velocity surveys constrains $\Omega_m$, or some combination of $\Omega_m$ and the galaxy bias parameter. Early studies in this field mostly assumed a pure matter universe and used $\gamma \approx 0.6$ (\citealt{peebles_peculiar_1976}), to either get bias independent measures of mass from velocity fields or to constrain $\Omega_m$ (see review articles by \citealt{dekel_dynamics_1994} and \citealt{strauss_density_1995}). The estimate by Peebles, given more than thirty five years ago, turned out to be a good approximation for a large range of models. 
The dependence of the linear growth rate on the cosmological constant (for e.g., \citealt{martel_linear_1991}; \citealt{lahav_dynamical_1991}) or on the dark energy equation of state $w$ (\citealt{wang_cluster_1998}; \citealt{linder_cosmic_2005}) was shown to be relatively weak. However, more recently, it has been pointed out that models with similar expansion histories but with different gravitational dynamics can be distinguished by their growth indices; $\gamma$ is 0.55 for the $\Lambda$CDM model, but 0.67 for certain modified theories of gravity (\citealt*{lue_probing_2004}; \citealt*{bueno_belloso_parametrization_2011}). 
Accordingly, modern observational efforts are focussed on measuring the linear growth rate with the added aim of constraining $\gamma$ (for e.g., \citealt{guzzo_test_2008}; \citealt{majerotto_probing_2012}). 

The method of velocity reconstruction from redshift surveys using linear theory breaks down when the perturbations are of order unity. Quasi-linear effects need to be included even to get an accurate determination of the linear growth rate (\citealt*{nusser_new_2012}). Various analytic quasi-linear and non-linear extensions using the Zel'dovich approximation (\citealt{nusser_cosmological_1991}; \citealt{gramann_improved_1993}), second order Eulerian perturbation theory (\citealt{bernardeau_quasi-gaussian_1992}, hereafter B92), and higher order Lagrangian perturbation theory (\citealt{gramann_second-order_1993}; \citealt{chodorowski_weakly_1997}; \citealt{chodorowski_recovery_1998}; \citealt{kitaura_estimating_2012}) have been proposed. Many of these analytic methods were also tested with numerical simulations (\citealt{mancinelli_nonlinear_1993}; \citealt{mancinelli_local_1995}). N-body codes, although accurate, give mass-weighted instead of volume-weighted estimates (\citealt*{dekel_potential_1990}) making it difficult to compare them with analytical answers. Refined volume averaged estimates were given using uniform-grid codes (\citealt{kudlicki_reconstructing_2000}; \citealt{ciecielg_gaussianity_2003}) and the method of tessellations (\citealt{bernardeau_new_1996}; \citealt{bernardeau_omega_1997}, \citealt{bernardeau_non-linearity_1999}). In general, these simulations are slow and are applicable to a limited range of models. 
It is usually assumed that the weak dark energy dependence of the linear relation extends to the non-linear regime and often the results are quoted in terms of the scaled velocity divergence $\nabla \cdot {\bf v}/f(\Omega_m)$, absorbing the cosmology dependence into the linear growth rate. While this assumption has been tested for different values of the cosmological constant $\Lambda$ (for e.g., \citealt{lahav_dynamical_1991}; \citealt{bouchet_perturbative_1995}; \citealt{nusser_omega_1998}), the explicit dependence on the dark energy equation of state $w$ has not yet been derived. 

The spherical top-hat system is an alternate way to model evolution in the non-linear regime. The model is simple to solve; equations of motion reduce to ordinary second order differential equations for the evolution of the scale factor (\citealt{Peebles80}). The main drawback is that it does not take into account interaction between scales; the price paid for computational ease. Nevertheless, it has been successful in predicting non-linear growth until virialization (\citealt{engineer_non-linear_2000}; \citealt{shaw_improved_2008}). Recently \cite{bilicki_velocity-density_2008}, hereafter BC08, obtained the non-linear density-velocity relation from the spherical top-hat. They utilised the top-hat's known exact analytic solutions and hence were restricted to pure matter cosmologies. We adopt a different approach, one that can be generalised to a range of background cosmologies. 
We numerically investigate how generic density and velocity perturbations evolve in a two dimensional density-velocity divergence phase space. We identify a special curve which is obtained by imposing the condition that the perturbations vanish at the big bang time and show that it acts like an attracting solution for the dynamics of the system. We refer to this as the `Zel'dovich curve' and demonstrate that it is the desired density-velocity relation in the non-linear regime. This approach was first put forward in a recent paper (\citealt{nadkarni-ghosh_extending_2011}, hereafter NC), but the analysis there was restricted to a EdS (with $\Omega_m =1$) cosmology. Here we extend the same idea to flat cosmological models with pressureless matter and non-clustering dark energy described by a constant equation of state $w$. 

The paper is organised as follows. \S \ref{sec:dynamics} gives the equations governing the spherical top hat, introduces the `Zel'dovich curve' and demonstrates its importance in the evolution of perturbations in the joint density-velocity phase space. \S \ref{sec:fits} gives a fit to the curve by generalising the formulae proposed by B92 and BC08 to include the dark energy term and this results in a new parametrization for the linear growth index $\gamma$ as a function of $w$. \S \ref{sec:cluster} discusses the evolution of galaxy cluster profiles using the spherical infall model and examines this curve in the context of observationally relevant quantities. \S \ref{sec:conc} gives the summary and conclusion. 

\section{Dynamics of the spherical top-hat}
\label{sec:dynamics}

\subsection{Physical set-up and equations }
\label{}
\def\ainit{{a_0}} 
\def\Hinit{{H_0}}
\def \rhomback{{\rho_m}}
\def \rhoXback{{\rho_{X}}}
\def \rhominitback{{\rho_{m,i}}}
\def \rhoXinitback{{\rho_{X,i}}}
\def \rhominitpert{{{\tilde \rho}_{m,i}}}
\def \rhoXinitpert{{{\tilde \rho}_{m,i}}}
\def \rhompert{{{\tilde \rho}_m}}
\def \rhoXpert{{{\tilde \rho}_X}}

\begin{figure}
\includegraphics[width=6cm]{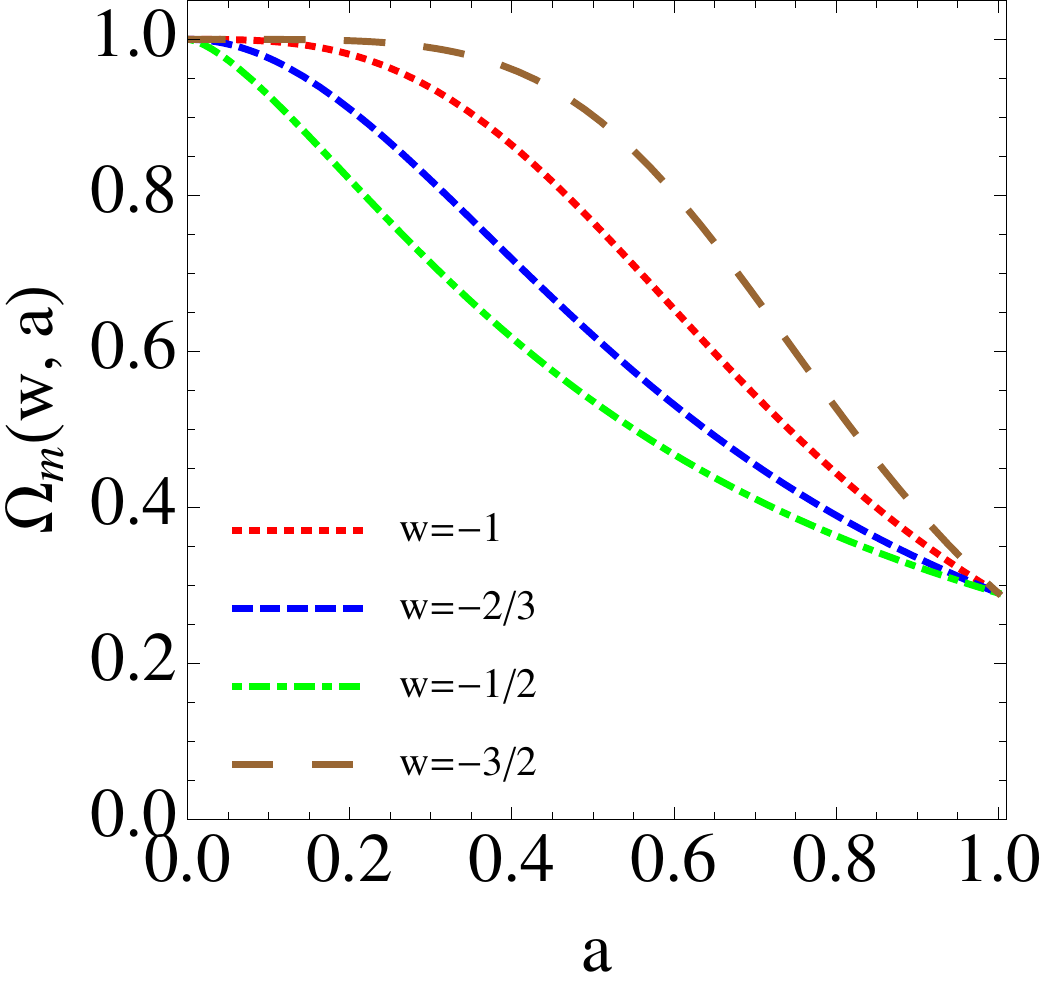}
\caption{Variation of $\Omega_m(a)$ for a flat cosmological model with dark matter and dark energy described by a constant equation of state $w$. For all models, $\Omega_{m,0} = 0.29$ and $\Omega_{\phi,0} = 0.71$. The values coincide at very early times and at the present epoch, but the evolution through intermediate epochs is different for different values of $w$.  }
\label{omegam}
\end{figure}
The compensated spherical top-hat perturbation consists of a uniform density sphere surrounded by a concentric spherical compensating region; homogenous and isotropic background extends beyond the outer edge of this region.  The compensating shell could be empty or contain mass depending upon whether the inner sphere is overdense or underdense with respect to the background. In this paper we restrict ourselves to flat background cosmologies comprising of pressureless matter and dark energy. Only matter perturbations are considered. The total matter density and Hubble parameter are denoted as $\rho_m({\tilde \rho}_m)$ and $H({\tilde H})$ for the background (perturbation). Dark energy is assumed to be spatially uniform and does not interact with the matter fields. It is described by a constant (in time) equation of state parameter $w$ defined as $ w= p_\phi/\rho_\phi$, where $p_\phi$ and $\rho_\phi$ are the dark energy pressure and density respectively.

The evolution of the background is completely determined by specifying $a_i$, $H_i$ $\rho_{m,i}$ and $\rho_{\phi,i}$ at some initial time $t_i$. The governing equation is 
\beq 
\frac{\ddot a}{a} = -\frac{H^2_i}{2} \left[\frac{\Omega_{m,i} a_i^3}{a^3} + (1+ 3 w) \Omega_{\phi,i} \left(\frac{a_i}{a}\right)^{3(1+w)}\right],
\label{backscalefac}
\eeq
where the dots represent derivatives with respect to time $t$ and $\Omega_{m(\phi),i}$ are the initial matter (dark energy) density parameters. The initial conditions for \eqnref{backscalefac} are $a(t_i) = a_i$ and ${\dot a}(t_i) = a_i H_i$.

The perturbation at the initial time is described by two quantities: the fractional overdensity $\delta_i$ and the fractional Hubble parameter $\delta_{v,i}$ defined as
\bea
\delta_i &=& \frac{\rhominitpert}{\rhominitback}-1,
\label{deltadefn} \\
\delta_{v,i} & = &\frac{{\tilde H}_i}{H_i}-1.
\label{deltavdefn}
\eea
In the case of a pure matter cosmology, Birkoff's theorem guarantees that the inner sphere can be described as a separate independent universe, whose scale factor obeys an equation analogous to \eqnref{backscalefac}, but with a different value for the matter density parameter. In the presence of a dark energy term the modification is not always obvious as has been discussed by various authors (\citealt*{pace_spherical_2010}; \citealt{wintergerst_clarifying_2010}). We follow the approach advocated in these papers of starting with the exact non-linear hydrodynamic equations for the density and velocity instead of an equation for the perturbation scale factor (for e.g., \citealt*{lima_newtonian_1997}; \citealt{abramo_structure_2007}). By recasting the Eulerian system in Lagrangian coordinates, one can show that for scales much smaller than the horizon scale and in the absence of matter-dark energy interactions, the spherical top-hat can be described by a scale factor $b(t)$ which obeys  
\beq
\frac{\ddot b}{b} = -\frac{H^2_i}{2} \left[\frac{\Omega_{m,i} a_i^3 (1+ \delta_i)}{b^3} + (1+ 3 w) \Omega_{\phi,i} \left(\frac{a_i}{a}\right)^{3(1+w)}\right],
\label{pertscalefac}   
\eeq
with initial conditions $b(t_i) = a_i$ and ${\dot b}(t_i) = {\dot a}_i (1+ \delta_{v,i})$ (see Appendix \ref{App:SCM} for details). In cosmology, the initial time is usually the time of recombination, after which perturbations start to grow. However, we will allow it to be specified anywhere between the big bang time (when $a=0$) and today (when $a=1$). $H_i$, $\Omega_{m,i}$, $\Omega_{\phi,i}$, $\delta_i$ and $\delta_{v,i}$ refer to the initial values at the specified initial epoch $a_i$.

The Hubble parameter $H$ and the density parameters ($\Omega_m, \Omega_\phi$) at any arbitrary epoch are related to those today ($H_0$, $\Omega_{m,0}, \Omega_{\phi,0}$) through
\bea
H^2(a) &=&H_0^2 \left[\frac{\Omega_{m,0}}{a^3} + \frac{\Omega_{\phi,0}}{a^{3(1+w)}} \right]
\label{eq:freid}\\
\Omega_m(a) &=& \frac{\Omega_{m,0} a^{-3}}{\Omega_{m,0} a^{-3} + \Omega_{\phi,0} a^{-3(1+w)}}, 
\label{omegamvar}\\
\Omega_\phi(a)   &=& \frac{\Omega_{\phi,0} a^{-3(1+w)}}{\Omega_{m,0} a^{-3} + \Omega_{\phi,0} a^{-3(1+w)}}.
\eea
The last two equalities follow from the definition of $\Omega$, the relation between $H$ and the critical density $\rho_c$ and the scalings $\rho_m \sim a^{-3}$, $\rho_\phi \sim a^{-3(1+w)}$. 
The evolution of $\Omega_m(a)$ is plotted in \figref{omegam} for four different values of the equation of state $w$. Flatness implies $\Omega_\phi(a) = 1-\Omega_m(a)$ at all epochs. $\Omega_{m,0}$ and $\Omega_{\phi,0}$ are fixed to be 0.29 and 0.71 in accordance with recent results from Type IA supernovae (SN Ia)  and {\it Wilkinson
Microwave Anisotropy Probe} (WMAP) data (\citealt{kowalski_improved_2008}, \citealt{komatsu_seven-year_2011}). From these results $w$ is known to be close to $-1$; we choose four values $w= -3/2, -1, -2/3, -1/2$. All models are matter dominated  at early epochs ($\Omega_m(a<<1) =1$) and have the same matter density $\Omega_{m,0} =0.29$ at late epochs. But at intermediate epochs, $\Omega_m(a)$ is different for the four models. Dark energy dominates earlier for larger values of $w$. For the EdS cosmology, $\Omega_m=1$ and $\Omega_\phi=0$ for all epochs. 

Once the solutions to \eqnrefs{backscalefac} and \eqnrefbare{pertscalefac} are known, the perturbation parameters at any later time can be computed 
\bea
\delta(t)  &=& \frac{(1+\delta_i) a^3}{b^3} -1,
\label{eq:delta}\\
\delta_v(t) &= &\frac{1}{H} \frac{{\dot b}}{b} -1. 
\label{eq:deltav}
\eea
The perturbation variable $\delta_v$ is related to the usual peculiar velocity divergence as 
\beq 
\delta_v = \frac{1}{3H } \nabla\cdot {\bf v}.
\eeq
In the rest of the paper we refer to $\delta$ as simply `density' and peculiar velocity as simply `velocity'.   

The spherical top-hat serves as a proxy for the non-linear regime only to the extent that it models density contrasts much higher than unity. It does not model non-linearities in truly inhomogeneous systems where different scales in the system interact to give non-local effects. In such cases the $\delta-\nabla \cdot {\bf v}$ relation has a scatter and can no longer be described by a one-to-one function. The `forward' relation (mean density in terms of the velocity divergence) and the `inverse' relation (mean velocity divergence in terms of the density) are not mathematical inverses of each other and it is more informative to describe the relation by 
a joint probability distribution function for the two variables (\citealt{bernardeau_non-linearity_1999}). These relations have been obtained for CDM and $\Lambda$CDM cosmologies using perturbation theory (e.g., B92; \citealt{chodorowski_weakly_1997}; \citealt{chodorowski_recovery_1998}) as well as simulations (e.g., \citealt{bernardeau_non-linearity_1999}; \citealt{kudlicki_reconstructing_2000}). The relation obtained from the top-hat is local and is expected to be applicable only in a average sense.

\subsection{The Zel'dovich curve}
\label{sec:zelcurve} 
\begin{figure*}
\includegraphics[width=17.5cm]{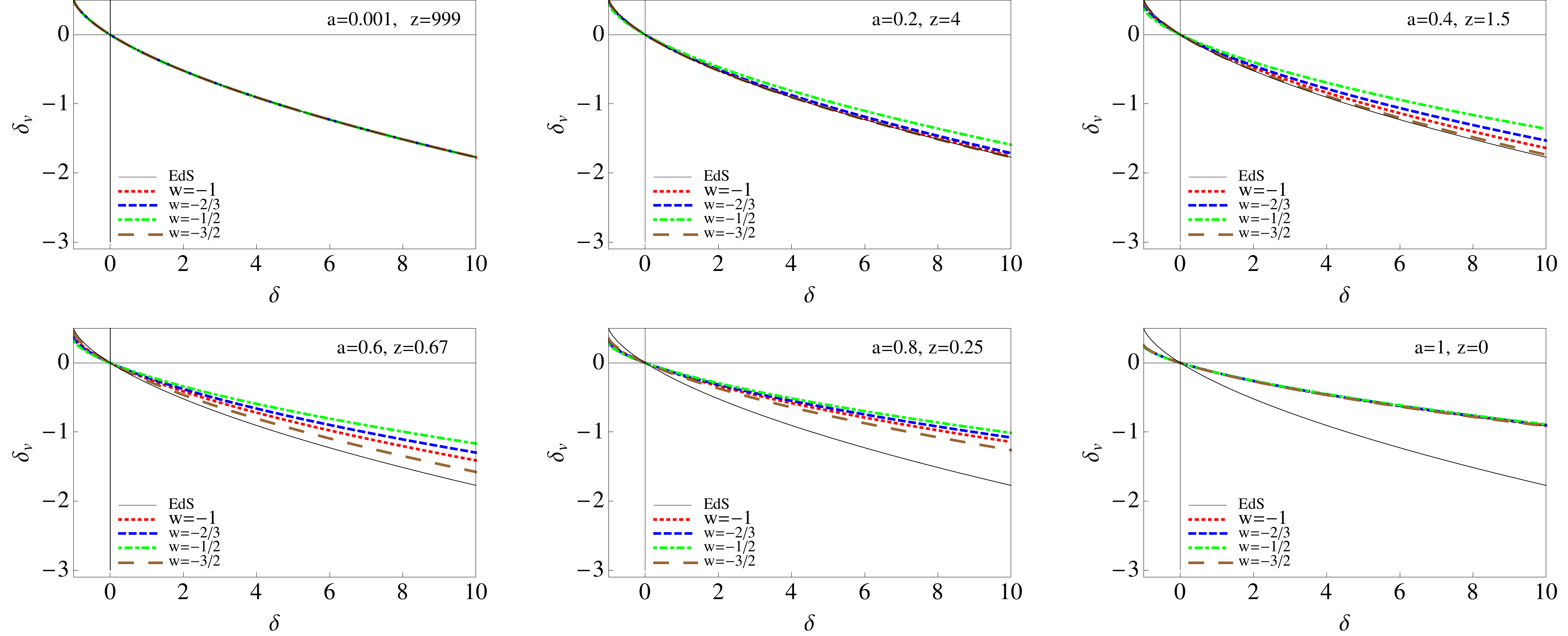}
\caption{Zel'dovich curves for different cosmologies at different epochs. Four dark energy models with $w=-3/2,-1, -2/3, -1/2$ are shown (bottom to top, colour code same as that in \figref{omegam}). The black curve is for the EdS case; it is the same at all epochs. In general, for a fixed value of $w$, the curves evolve with time. At early times, the various curves coincide with the EdS curve and almost coincide with each other at the present epoch, deviating at intermediate epochs. This indicates that they are mainly dependent on $\Omega_m$ and the explicit dependence on $w$ is very weak. }
\label{fig:zeldovich}
\end{figure*}
Consider the two-dimensional phase space whose abscissa and ordinate are $\delta$ and $\delta_v$ respectively. Mathematically, the initial value of $\delta$ and $\delta_v$ are independent choices and this freedom allows for solutions where the perturbation scale factor is non-zero at the big bang time. However, physically, it is reasonable to expect that there were no perturbations at the big bang epoch i.e., the scale factors of the background and the perturbation are 
both zero at the big bang and start evolving since then. This condition sets a specific relationship between the initial $\delta$ and $\delta_v$. If we define the age at time $t_i$ to be the time taken for the scale factor to grow from zero to its value at $t_i$ 
 then this condition is equivalent to demanding that the ages of the background and perturbation be the same. 
For any starting epoch, this relation traces out a curve in the $\delta-\delta_v$ phase space. The `equal age' condition is imposed in linear theory by ignoring the decaying modes in the solution for $\delta$. This assumption also forms the basis of the Zel'dovich approximation (\citealt{zeldovich_gravitational_1970}) and hence we refer to this special curve as the `Zel'dovich curve'.

In NC this curve was examined for the EdS cosmology. In this special case, from \eqnrefs{backscalefac} and \eqnrefbare{pertscalefac}, one can explicitly calculate the ages of the background and perturbation and the `equal age' condition becomes (see Appendix \ref{App:zellcdm})
\beq
\int_{y=0}^{y=1}\frac{dy}{\left[(1+\delta) y^{-1} +  (1+ \delta_v)^2 - (1+\delta) \right]^ {1/2}} = \frac{2}{3}.
\label{eq:omegam1}
\eeq
This relation does not involve any time dependent quantities and the Zel'dovich curve has the same form at all epochs. 
However, in the presence of generic dark energy terms, this relation has a time dependence. This can be seen by scaling $a$ and $b$ in \eqnrefs{backscalefac} and \eqnrefbare{pertscalefac} by $a_i$ and replacing time by $t H_i$. The equations and initial conditions depend only on the parameters $\delta_i, \delta_{v,i}, \Omega_{m,i}$ and $w$ (flatness removes $\Omega_{\phi,i}$ dependence). The `equal age' criterion remains unchanged by this scaling and the resulting $\delta_i-\delta_{v,i}$ relation depends only on the parameters $\Omega_{m,i}$ and $w$. 
Since the value of $\Omega_{m,i}$ changes with the starting epoch, the relation will be different at different epochs. For the kind of dark energy models considered here, analytic solutions for 
\eqnrefs{backscalefac} and \eqnrefbare{pertscalefac} are somewhat tedious, if at all possible (\citealt{lee_spherical_2010}), and we solve the equations numerically. Appendix \ref{App:zellcdm} gives the details of the calculation. 

\capfigref{fig:zeldovich} shows the Zel'dovich curves for different cosmological models. Brown (long dashed), red (dotted), blue (short dashed) and green (dot dashed) lines correspond to $\Omega_{m,0} =0.29, \Omega_{\phi,0} =0.71$ and $w=-3/2, -1, -2/3, -1/2$ respectively. The black curve, given by \eqnref{eq:omegam1}, corresponds to the EdS model. Six different epochs between $a=0.001$ and $a=1$ or, equivalently, between $z=999$ and $z=0$ are considered. At recombination ($a \approx 0.001$), the curves for different $w$ overlap with each other and with the EdS case. At $a=1$ they again overlap, but differ from the EdS case. At intermediate epochs, they differ from each other. Comparison with \figref{omegam} suggests that the Zel'dovich curve mainly depends on $\Omega_m$ and the explicit dependence on $w$ is very weak. 

In the next section we will show that these curves play a special role in the dynamics of the perturbations in the $\delta-\delta_v$ phase space.

\subsection{Dynamics in the $\delta-\delta_v$ phase space}
\label{subsec:dynamics}
\begin{figure*}
\includegraphics[width=17cm]{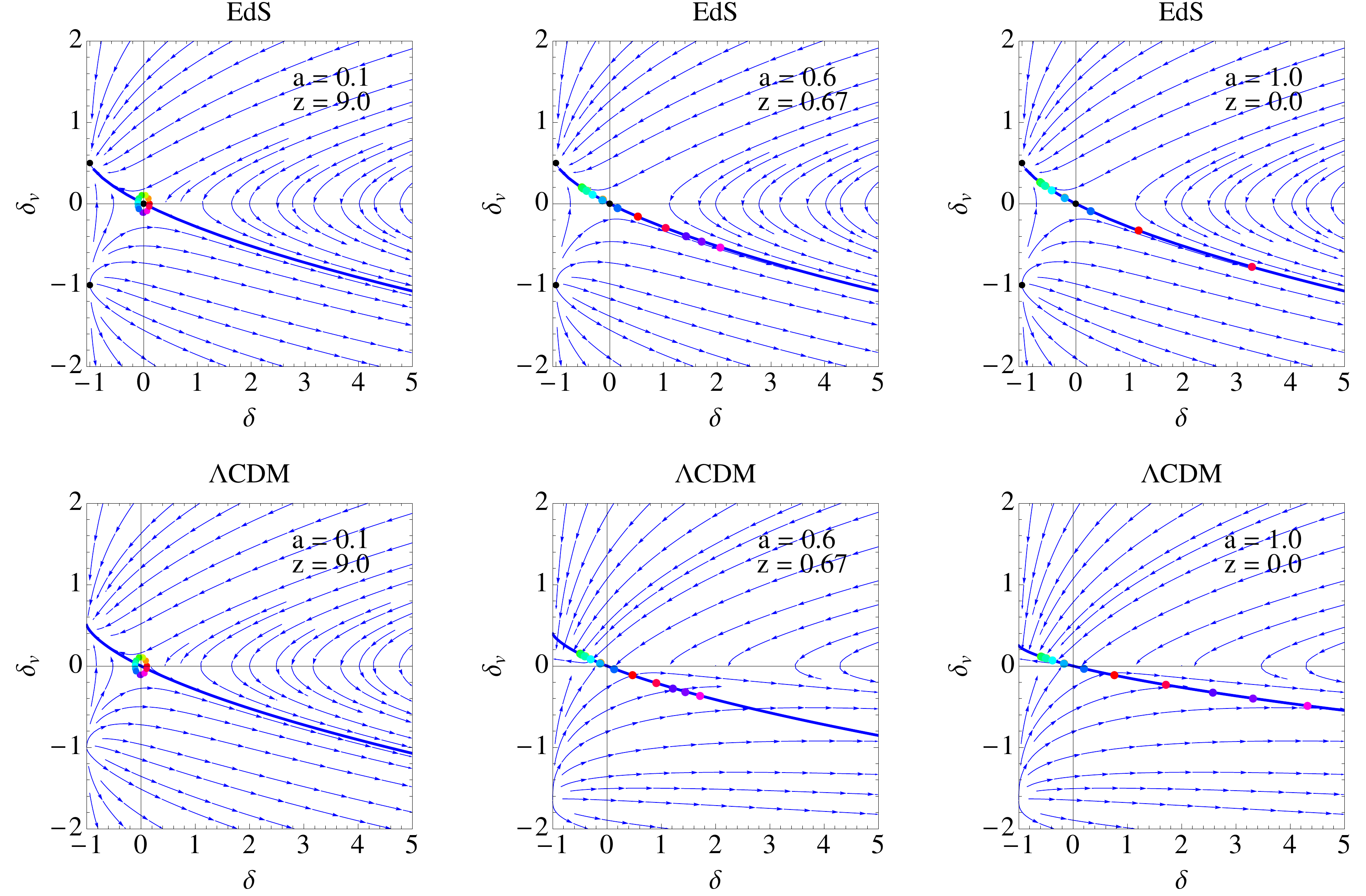}
\caption{Phase space flow for two different cosmological models. The upper panel is the EdS model; the Zel'dovich curve stays the same at all times and is an attractor of the flow. The lower panel is the $\Lambda$CDM model with $\Omega_{m,0}=0.29$, $\Omega_{\phi,0}=0.71, w=-1$. The Zel'dovich curve is different at different times and so is the instantaneous phase portrait, a signature of the non-autonomous nature of the system. The coloured ring of initial perturbations (coloured dots) starts near the origin at $a=0.1$. As the evolution proceeds the perturbations asymptotically approach the Zel'dovich curve (numerical values in text). The curve acts like an attractor and is the required non-linear extension of the density-velocity divergence relation. } 
\label{fig:flowpattern}
\end{figure*}
The definitions in \eqnrefs{eq:delta} and \eqnrefbare{eq:deltav} combined with \eqnrefs{backscalefac} and \eqnrefbare{pertscalefac} give equations that govern the phase space evolution of $\delta$ and $\delta_v$:
\bea
{\dot \delta} &=& -3 H \delta_v (1+ \delta) 
\label{DVDRdyn1}
\\
{\dot \delta_v} &=& - \frac{H}{2} \left[\Omega_m(a)\delta - \delta_v \{\Omega_m(a) +(1+3w) \Omega_\phi(a) -2\} + 2\delta_v^2 \right ].
\label{DVDRdyn2}
\eea
In general, because of the time variation of $\Omega_m(a)$, the system defined by \eqnrefs{DVDRdyn1} and \eqnrefbare{DVDRdyn2} is non-autonomous for dark energy models and techniques of linear stability analysis that are usually performed on autonomous systems are not applicable. However, it is instructive to first consider the EdS case for which the temporal dependence drops and the system becomes autonomous. \capfigref{fig:flowpattern} shows the phase portrait for the two qualitatively different cases: EdS (upper panel) and $\Lambda$CDM (lower panel) at three different epochs $a=0.1,0.6,1$ corresponding to redshifts $z=9, 0.67, 0$ respectively. The blue lines with arrows are the streamlines of the flow. A streamline is drawn such that the tangent at any point gives the direction of the phase space velocity vector $({\dot \delta}, {\dot \delta}_v)$ at that point. It denotes the instantaneous direction of evolution of the system. For an autonomous system, the velocity vector is independent of time and the pattern of streamlines is the same at all epochs. However, for a non-autonomous system, the phase space velocity has a time dependence and hence the pattern of streamlines changes with $a$.  

For the EdS case, the system has three fixed points, shown by the black dots, at $(0,0), (-1,-1)$ and $(-1,0.5)$. The saddle point at the origin $(0,0)$ corresponds to a unperturbed cosmology, the unstable node at $(-1,-1)$ corresponds to a vacuum static model and the attracting point at $(-1,0.5)$ corresponds to a expanding void model. The solid blue curve is the Zel'dovich curve. The multi-coloured ring of points at $a=0.1$ corresponds to initial conditions with different values of $\delta$ and $\delta_v$, all at a distance of $0.11$ from the centre. Some are very close to the Zel'dovich curve at the start, but others are off the curve. However, as the flow proceeds, all the points evolve to lie close to the Zel'dovich curve at $a=0.6$ and $a=1$. 
The relative deviations from the curve are estimated as $|\delta_{ evol.}(\delta_v)/\delta_{Zel.}(\delta_v) -1|$, where at a given value of $\delta_v$, $\delta_{evol.}$ is the value evolved from initial conditions at $a=0.1$ and $\delta_{Zel.}$ is the value given by the definition of the Zel'dovich curve. The closeness to the curve is characterised by the maximum relative deviation over the points considered for the evolution. Thirteen points were considered and the maximum relative deviation was 14\% at $a=0.6$ and 3\% at $a=1$. The deviation decreases with time indicating that the trajectories asymptotically approach the curve. A point that starts along the curve at $a=0.1$ was found to remain along the curve with a maximum relative deviation of $10^{-8}$ over the entire range of evolution from $a=0.1$ to $a=1$. Thus, the Zel'dovich curve forms an invariant set and acts like an attracting solution for the system.

For the $\Lambda$CDM case, at $a=0.1$, $\Omega_m \approx 1$ and $\Omega_\phi \approx 0$. Hence the flow pattern resembles that of the EdS cosmology and the Zel'dovich curves of the two cases are coincident. But as was discussed in the previous section, the Zel'dovich curve evolves with time.  It is possible to discuss the stability of such non-autonomous flows in more formal terms using tools from dynamical systems theory (for e.g. \citealt*{shadden_definition_2005}), but for the purposes of this paper, we will simply illustrate the `attracting' behaviour of the Zel'dovich curve. We consider the same set of initial conditions, represented by the multi-coloured ring $a=0.1$, and evolve them using \eqnrefs{DVDRdyn1} and \eqnrefbare{DVDRdyn2}. At $a=0.6$ and at $a=1$ the set has evolved to lie close the corresponding Zel'dovich curves at that epoch. The maximum relative deviations from the corresponding Zel'dovich curve were defined similar to the  earlier case, but at fixed $\delta$ instead of $\delta_v$. The maximum relative deviation was 17\% at $a=0.6$ and 6\% at $a=1$, again indicating that the Zel'dovich curve acts like an attractor. A point that starts along the curve at $a=0.1$ stays along it with a maximum relative deviation of $10^{-6}$ over the entire range of evolution. Hence, even in the $\Lambda$CDM case, initial conditions that satisfy the Zel'dovich relation at the start continue to maintain it and those that do not, evolve such that they establish it. 

Thus, the Zel'dovich curve has a special significance in the dynamics of the perturbations as they evolve in phase space. It gives the long term behaviour of the density and velocity perturbations and is the exact non-linear extension of the density-velocity divergence relation for the spherical top-hat. In the next section we will see that a generalisation of existing formulae by B92 and BCO8 provide a 3\% accurate fit to this curve. In \figref{fig:flowpattern} the deviations from the Zel'dovich curve of the evolved points was greater than this accuracy and one may argue that the fitting forms cannot be useful approximation of the dynamics. However, the initial values of $a=0.1$ and the ring of radius 0.11 were chosen merely for the sake of convenient plotting and to demonstrate the asymptotic behaviour. Real cosmological initial conditions start evolving at recombination ($a=0.001$) and sense the presence of the Zel'dovich attractor for a longer period of time. We evolved a similar ring of initial points with an amplitude of $1.1\times 10^{-3}$ from $a=0.001$ to $a=1$. The maximum relative deviation over the entire range was within 0.01\% indicating that the fits in the following section are indeed a good approximation to the dynamics at the percent level.

\section{Fitting functions and growth rates}
\label{sec:fits}
In this section we show that appropriate modifications of already existing formulae provide a good fit for the Zel'dovich curves. The modification effectively changes the linear growth rate to account for the inclusion of dark energy.  Using these fits we also obtain an approximate formula for the growth rate in the non-linear regime. 

\subsection{Generalisation of existing forms}
\begin{figure*}
\includegraphics[width=16cm]{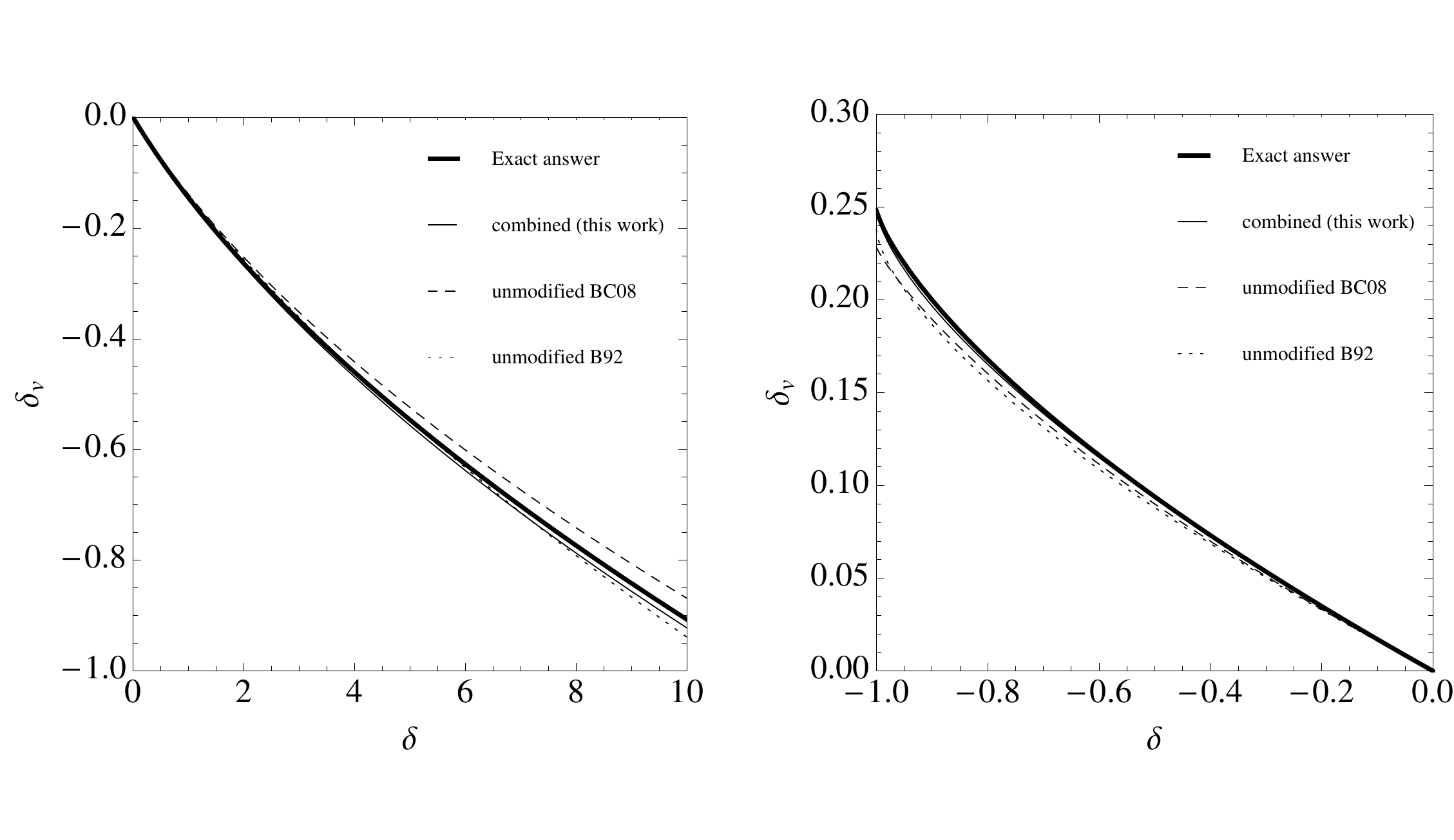}
\caption{Comparison of the various approximations of the density-velocity relationship in literature for the $\Lambda$CDM model at $a=1$. The left (right) panels show the relationship for overdensities (voids). The combined 3\% fit is a modified version of B92 (\eqnref{eq:B92}) over the range $-1\leq \delta \leq 1$ and of BC08 (\eqnref{eq:BC08}) over the range $1<\delta \leq 10$ and does marginally better than the unmodified formulae which were derived for pure matter models. The values $\Omega_{m,0} =0.29$ and $\Omega_{\phi,0}=0.71$ were used for the comparison.}  
\label{fig:compare}
\end{figure*}

In the past, various formulae have been proposed for the non-linear density-velocity divergence relation and recent paper by \cite{kitaura_estimating_2012} gives a nice summary. Amongst them, the fit by B92 is perhaps the simplest because it has the fewest parameters. However, BC08 showed that this formula was inaccurate for empty universes and provided an improved version based on analytic results of the spherical top-hat model. While their fit has a simple form for overdensities, it has a more involved form for voids. We found that a simple combination of the two formulae by B92 and BC08, appropriately modified to account for the $w$ dependence provides a good fit for the Zel'dovich curves.

The original \footnote{Note that the definition of $\delta_v$ in B92 and our $\delta_v$ differ by a factor of 3; 
$\delta_{v, B92} = 3 \delta_{v}$. Similarly, the variable $\Theta$ of BC08 is related to our $\delta_v$ as $\Theta = -3 f^{-1}(\Omega_m)\delta_v$. The formulae quoted here have taken these differences into account so that a direct comparison is possible.} B92 formula  is 
\beq 
\label{eq:B92}
\delta_v = \frac{\Omega_m^{0.6}}{2} [1-(1+\delta)^{2/3}].
\eeq
This formula is expected to be valid in the range $-1\leq \delta\leq 2$. The original BC08 formula is
\beq
\label{eq:BC08}
\delta_v = \left \{\begin{array}{cc}
f(\Omega_m)[(1+\delta)^{1/6}-(1+\delta)^{1/2} ] & \delta \geq 0\\
-\frac{f(\Omega_m)}{3}[\delta +(1+ \Theta_{min}(\Omega_m)) N(\delta) ] &-1\leq\delta <0
\end{array}
\right.
\eeq
where 
\bea 
f(\Omega_m) &\simeq & \Omega_m^{0.6}\\
\Theta_{min}(\Omega_m) & \simeq & -1 -0.5 \Omega_m^{0.12 -0.6 \Omega_m},\\
N(\delta) &= & (\delta+1) \ln(\delta+1) -\delta.
\eea
Both the B92 and the BC08 formulae are obtained for pure matter cosmologies. As was shown in \S \ref{sec:zelcurve} there is also expected to be a weak $w$ dependence. For a given $w$, $\Omega_m$ varies with $a$ and this corresponds to a family of curves for each value of $w$. 
To estimate the joint dependence on $\Omega_m$ and $w$, Zel'dovich curves were obtained for seven different values of $w$ in the range $-1/2$ to $-3/2$ and at eleven epochs in the range $a=0.001$ to $a=1$. The $\delta$ range for each curve was restricted to $-1\leq \delta \leq 10$. In the range $-1\leq \delta \leq 1$ we chose a generalised form of the B92 formula

\beq
\label{eq:B92mod} \delta_v = A(\Omega_m, w) (1 - (1+\delta)^{B(\Omega_m,w)}).
\eeq
The numerical $\delta=-1$ limit of the Zel'dovich curve and the slope of the curve in the regime $|\delta|<0.01$ fixed the two unknowns $A(\Omega_m, w)$ and $B(\Omega_m,w)$.  
\bea
A(\Omega_m,w) &=& \frac{\Omega_m^{\gamma_1(w)}}{2}
\label{eq:A} \\ 
B(\Omega_m,w) &=& \frac{2}{3}\Omega_m^{\gamma_2(w)}, 
\label{eq:B}
\eea
where 
\bea 
\label{gamma1}\gamma_1(w) &=& 0.56 (-w)^{-0.08}\\
 \gamma_2(w) &=& -0.01 (-w)^{-1.18}.
 \label{gamma2}
 \eea 
This fit gave a maximum relative error of $3 \% $ within the range of the fit. It gave a $11\%$ error in a higher range of $\delta$ values $1 \leq \delta \leq 10$. In this range, a modification of the formula of BC08 for positive $\delta$ provides a better fit with $3\%$ maximum relative error 
\beq
\label{eq:BC08mod} \delta_v = \Omega_m^{\gamma_1(w) + \gamma_2(w)} ((1+\delta)^{1/6} - (1+\delta)^{1/2}),
\eeq
with the same $\gamma_1(w)$ and $\gamma_2(w)$ defined in \eqnrefs{gamma1} and \eqnrefbare{gamma2}.

The linear limit gives 
\beq 
\delta_v = -\frac{f(\Omega_m) \delta}{3}, 
\eeq
 with 
\beq 
f(\Omega_m,w) = \Omega_m^{\gamma_1(w) + \gamma_2(w)}.
\eeq
The linear growth index is $\gamma(w) = \gamma_1 + \gamma_2 = 0.56(-w)^{-0.08} -0.01 (-w)^{-1.18}$. 
We compare this new parametrization for the growth index to the widely accepted result of \citet{linder_cosmic_2005} 
\beq
\gamma(w) = \left\{
\begin{array}{cc}
0.55+0.05[1+w(z=1)] & w \geq -1\\
0.55+0.02[1+w(z=1)] & w <-1.
\end{array}
\right.
\eeq
For a $\Lambda$CDM cosmology Linder's value of $\gamma=0.55$ is recovered exactly. For other values in the range $-3/2<w<-1/2$, the relative differences are at most 1\%. This fit also modifies the void limit ($\delta=-1$) from $\delta_v = \frac{\Omega_m^{0.6}}{2}$ for the B92 formula to $\delta_v = \frac{\Omega_{m}^{0.56(-w)^{0.08}}}{2}$. This limit is often important when fits to simulation results are compared to analytical estimates (for e.g., \citealt{bernardeau_omega_1997}; \citealt{kudlicki_reconstructing_2000}).  Note that for most reasonable values of $w$, $\gamma_2$ is rather small and the $\delta$ dependence of the B92 formula is almost unchanged. Similarly, the only change to the overdensity formula of BC08 is in the linear growth factor $f(\Omega_m)$.

\capfigref{fig:compare} compares the unmodified forms of B92 and BC08 (\eqnrefs{eq:B92} and \eqnrefbare{eq:BC08}) with the fit presented in this paper for the $\Lambda$CDM case at $a=1$. For this comparison, $\Omega_{m,0}=0.29, \Omega_{\phi,0}=0.71$. The left panel compares positive overdensities, the right panel compares voids. Contrary to what BC08 presented in their paper (fig. 5 in BC08), the {\it unmodified} B92 fit seems to perform better than the {\it unmodified} BC08 even for higher values of $\delta$. This could be because the value of $\Omega_m$ chosen for the comparison is different ($\Omega_{m,0} =0.25$ in BC08 vs. $\Omega_{m,0} =0.29$ here) and the comparison there was between matter models as opposed to the $\Lambda$CDM model here. This contradiction is somewhat resolved when one compares the modified versions. As mentioned earlier, when compared over the entire range of $w$ ($-3/2 \leq w\leq -1/2$) and epochs ($0.001\leq a\leq1$), the {\it modified} version of BC08 performed better (3\% max. relative error) at higher $\delta$ than the {\it modified} B92 (11\% max. relative error) and we advocate using the former for the high $\delta$ regime. 

How much improvement has been brought about by the modified combined fit ? We found that over the entire range of $w$ and $a$ the unmodified B92 (BC08) fits are about 11\% (10\%) accurate whereas the modified, combined version is 3\% (marginally better). 
In this paper, we have assumed spherical symmetry to derive the non-linear $\delta-\delta_v$ relation. As was commented in BC08, better agreement with the spherical system does not guarantee better agreement with the real system. Numerical simulations or other non-linear techniques will be required to give more refined estimates of this relation, nevertheless, the Zel'dovich curve obtained for the spherical system can provide a starting point to analyse the results.

\subsection{Non-linear growth rate}
\label{sec:nonlingrowth}
\begin{figure*} 
\includegraphics[width=16.cm]{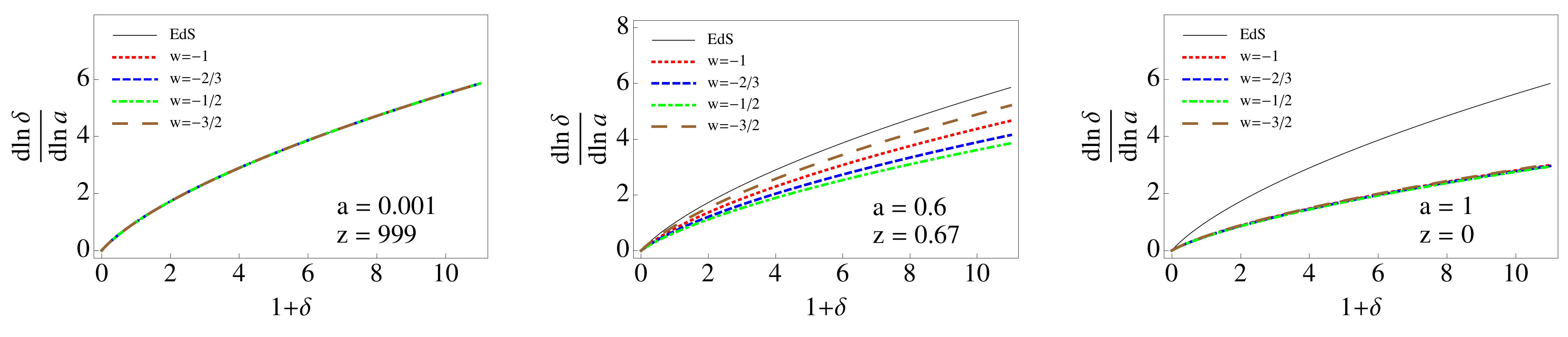}
\caption{Non-linear growth factor as a function of $(1+ \delta)$ for models with four different values of $w$ and the EdS model. Colour coding is same as in \figref{fig:zeldovich}. Growth rates for the various models deviate only at intermediate epochs because the values of $\Omega_m$ are different in the different models. Lower value of $w$ implies higher value of $\Omega_m$ and hence a higher growth rate. }
\label{fig:growthrate}
\end{figure*}
The Zel'dovich curve gives the instantaneous location of the perturbations in phase space. It does not directly give information about  the rate at which perturbations evolve along the curve. This is encoded in the {\it growth rate}, usually defined as $d{\rm ln} \delta/d{\rm ln} a$. From \eqnref{DVDRdyn1}, we see that the growth rate for the spherical system can be expressed as
\beq 
f = \frac{d {\rm ln} \delta}{d {\rm ln} a} = - \frac{3 \delta_v (1+\delta)}{\delta}.
\label{nonlinearf}
\eeq
This expression is valid both in the linear and non-linear regimes and we refer to $f$ as the total growth rate or non-linear growth rate. Along the Zel'dovich curve, $\delta_v$ is a function of $\delta$ and the growth rate can be expressed solely as a function of $\delta$. \capfigref{fig:growthrate} shows the growth rate vs. $(1+\delta)$ along the Zel'dovich curve for different values of $w$ at three different redshifts. The colour coding is same as that in \figref{fig:zeldovich}. $(1+\delta)$ is a more natural variable than $\delta$ since the former is a measure of the total density of the system. As expected, the growth rate is higher for higher values of $1+\delta$; larger densities grow faster, a signature of gravitational instability. At intermediate epochs, the growth rate is higher for lower values of $w$ because the values of $\Omega_m$ are higher (see \figref{omegam}). The cumulative effect of higher growth factors can be seen by referring back to \figref{fig:flowpattern}. Consider one of the collapsing perturbations, for example the one shown by the red dot in the upper and lower panels. It starts at the same point in the phase space at $a=0.1$ for both models. At $a=1$, it has evolved further along the Zel'dovich curve for the EdS case as opposed to the $\Lambda$CDM case. The perturbations shown by the purple and violet dots are already out of the plot at $a=1$ in the upper panel whereas are still at $\delta<5$ in the lower panel. 

Using the fits above, the non-linear growth rate in the regime $-1\leq \delta\leq 1$ is 
\beq 
f(\Omega_m, w, \delta) = -\frac{3 A(\Omega_m,w)\left[(1+\delta) -(1+\delta)^{B(\Omega_m,w)+1}\right] }{\delta}. 
\label{non-linfit}
\eeq
Substituting for $A$ and $B$ using \eqnrefs{eq:A} and \eqnrefbare{eq:B} and in the limit of small $\delta$, we get 
\beq 
f(\Omega_m, w, \delta) \approx \Omega_m^{\gamma_1 + \gamma_2} \left[ 1+ \frac{(\frac{2}{3} \Omega_m^{\gamma_2} +1) \delta}{2} + \mathcal{O}(\delta^2)\right] .\label{non-linfit2}
\eeq
To lowest order this reduces to the linear growth rate $f_{lin} = \Omega_m^{\gamma_1 + \gamma_2}  $. At the next order, the non-linear correction is linear in $\delta$. It would be interesting to check if numerical simulations show a similar $\delta$ dependence of the non-linear growth factor. 

\section{Connection to observables: Galaxy cluster profiles }
\label{sec:cluster}

Studies of the matter density distribution and pattern of infall velocities around a rich cluster of galaxies have traditionally been used to provide constraints on the matter density $\Omega_m$ (\citealt{gunn_infall_1972}; \citealt{peebles_peculiar_1976}; \citealt{regos_infall_1989}; \citealt{willick_maximum_1997}; \citealt{willick_maximum_1998}). More recently, galaxy cluster profiles have also been used as a test of modified gravity (\citealt{lombriser_cluster_2012}). Here we investigate dependence of the cluster density and velocity profiles on the dark energy equation of state. We use the spherical infall model (e.g., \citealt{silk_primordial_1979-1}; \citealt{silk_primordial_1979}; \citealt{villumsen_velocity_1986}; \citealt{lilje_evolution_1991}, hereafter LL91) to track the evolution of the cluster into the mildly non-linear regime. Our overall set-up is similar to that described in \cite{lahav_dynamical_1991}.

\subsection{Equations for spherical infall}
\label{subsec:eq}
The initial configuration of the system is similar to that of the top-hat, but with a radial density variation and hence the perturbation cannot be described by a purely time dependent scale factor. The detailed description of the system and derivation of the relevant equations is given in Appendix \ref{App:SCM}. Here we state the main results. 
The system is described by a continuum of shells with Lagrangian coordinates $X=r(t_i)/a_i$, where $r(t_i)$ is the initial physical radius of the shell. The radius of the shell at any later time 
\beq
r(X, t) = b(X,t) X.
\label{eq:r}
\eeq
The initial perturbation is completely specified by its initial density and velocity profiles. \capeqnrefs{deltadefn} and \eqnrefbare{deltavdefn} generalise to
\bea
\nonumber \delta(X, t_i)& = & \frac{{\tilde \rho}_m(X,t_i)}{\rho_m(t_i)} -1, \\
\nonumber \delta_v(X, t_i) & = & = \frac{1}{H_i} \frac{{\dot r}(X,t_i)}{r(X,t_i)} -1
\eea 
and the evolution of $b(X,t)$ is given by 
\beq
\frac{\ddot b}{b} = -\frac{H_i^2}{2} \left[\frac{ \Omega_{m,i} (1+ \Delta(X,t_i)) a_i^3 }{b^3}  + (1+ 3 w) \Omega_{\phi,i} \left(\frac{a_i}{a}\right)^{3(1+w)} \right],
\label{eq:clusterevol}
\eeq
with initial conditions $b(X,t_i) = a_i$, ${\dot b}(X,t_i) = {\dot a}_i (1+ \delta_v(X,t_i))$ and 
\beq 
\Delta(X, t_i) = \frac{3}{X^3} \int_0^X  \delta(X', t_i)  X'^2 dX',
\label{eq:Deltainit}
\eeq
where the spherically averaged density inside any general radius `r' is given by 
\beq
\Delta(r) = \frac{3}{r^3} \int_0^r \delta(r') r'^2  dr'.
\label{eq:Deltadef}
\eeq
The Jacobian that relates the initial Lagrangian and the physical coordinates is 
\beq
J(X, t) = \frac{r^2}{X^2} \frac{dr}{dX} = b^3 \left(1+ \frac{X}{b} \frac{db}{dX} \right) 
\label{eq:jacob} 
\eeq
and the density at any later time is
\beq
 \delta(X, t) = \frac{(1+ \delta(X, t_i))a^3}{J(X,t)} -1 = \frac{(1+ \delta(X, t_i))a^3}{b^3 \left(1+ \frac{X}{b} \frac{db}{dX} \right)}-1.
\label{eq:deltaevol}
\eeq
Note that the evolution of $\delta$ is different from that given by \eqnref{eq:delta} due to the spatial dependence of the perturbation scale factor $b(X,t)$ and the combination $\delta-\delta_v$ no longer satisfies the Zel'dovich relation. Instead the relevant quantity is the average density $\Delta$. 
Substituting \eqnref{eq:r} in \eqnref{eq:Deltadef} and using the relations \eqnrefs{eq:Deltainit}, \eqnrefbare{eq:jacob} and \eqnrefbare{eq:deltaevol} gives 
\beq
\Delta(X, t) =  \frac{(1+\Delta(X, t_i)) a^3}{b^3} -1.  
\label{eq:Deltaevol}
\eeq
Comparing \eqnrefs{eq:clusterevol} and \eqnrefbare{eq:Deltaevol} with \eqnrefs{pertscalefac} and \eqnrefbare{eq:delta}, it is clear that the role played by $\delta$ in the uniform density case is now played by $\Delta$. Physically, this is expected because the velocity is set by the net mass inside the shell. The velocity divergence parameter is similar to \eqnref{eq:deltav}, but with a spatial dependence
\beq 
\delta_v(X,t) = \frac{1}{H} \frac{{\dot b}(X,t)}{b(X,t)}-1.
\eeq
The Zel'dovich curve becomes a $\Delta-\delta_v$ relation. 

\subsection{Initial conditions}
The initial density profile $\delta(X, t_i)$ was chosen according to the prescription described in LL91 with some minor modifications. This prescription uses the theoretical framework of \cite{bardeen_statistics_1986} to predict the density profile around a primordial density peak that nucleates a cluster. The cosmological parameters today were set as $\Omega_{m,0}=0.29$, $\Omega_{\phi,0} = 0.71$, $H_0 =  100 h$ km${\rm s}^{-1} {\rm Mpc}^{-1}$. The r.m.s. fluctuation on $8 h^{-1}{\rm Mpc}$, $\sigma_8$, fixed the normalisation of the initial power spectrum. We ran two sets of runs with $\sigma_8 =0.8$ and $0.9$. In each set, three equation of state parameters were considered: $w= -1, -1/2, -3/2 $. The linear density profile today was chosen to be the same for all three cases and the initial profiles at $z=1000$ are obtained by multiplying by the appropriate linear growth rate factor. Therefore, the amplitude of the initial density profile was different for the three cases. The initial velocity profile was determined by imposing the linear Zel'dovich condition in the $\Delta-\delta_v$ phase space: $\delta_v(X, t_i) =  -1/3 \Omega_m^{\gamma_1+\gamma_2} \Delta(X, t_i)$.

 \capeqnref{eq:clusterevol} for $b(X,t)$ was integrated numerically for a discrete set of shells from the initial time $z_i = 1000$ to the final time $z_f=0$. During the course of evolution if $b=0$, the shell is considered to have collapsed, and forms a part of the cluster core \footnote{Ideally, for collisionless infall, a shell that collapses, bounces back and forth before eventually settling down to a periodic bouncing motion with maximum radius equal to 0.8 times its turnaround radius (\citealt{bertschinger_self-similar_1985-1}). In this case, the shell will collide with outer shells that are undergoing their initial collapse and the equations of motion will be differ from \eqnref{eq:clusterevol}. In this paper we follow \cite{lahav_dynamical_1991} and ignore these effects of secondary infall.}. The initial set up is such that the inner shells collapse before the outer shells and no shell crossing takes place anywhere other than at the centre. Further details regarding the initial set-up can be found in Appendix \ref{app:initden}. In the entire discussion, we assume that light traces matter exactly i.e., the bias factor between galaxies and dark matter is unity. The final density profile is computed using \eqnref{eq:deltaevol} and the final infall velocity is 
\beq
v_{infall}(X,t) = H r(X,t) - {\dot r}(X,t).
\eeq
$\delta_v$ and $v_{infall}$ are related as 
\beq 
\delta_v(X,t) = -\frac{v_{infall}(X,t)}{H r(X,t)}.
\eeq

\subsection{Results}

\begin{figure*}
\includegraphics[width=16cm]{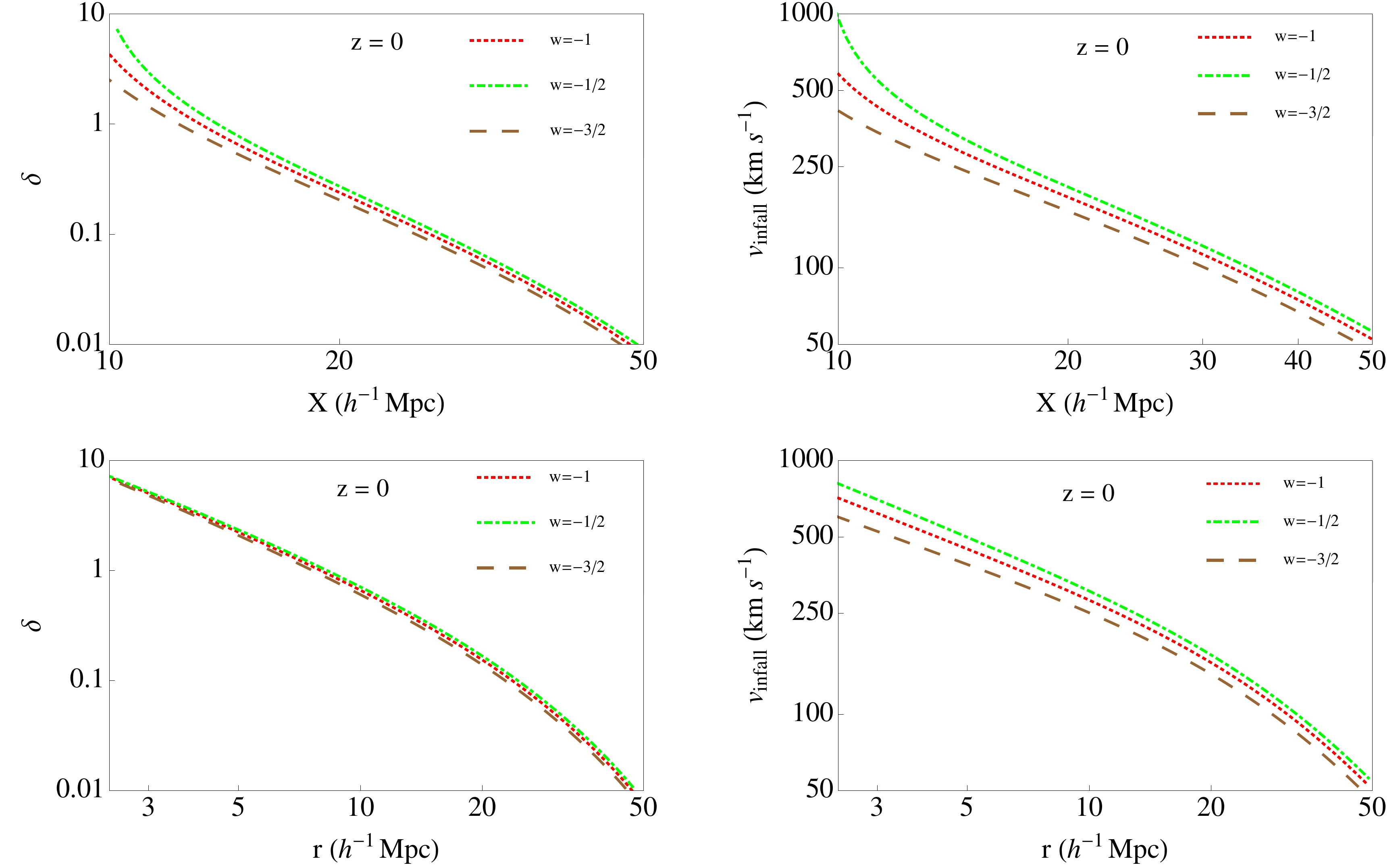}
\caption{The density and infall velocity profiles vs. $X$, the initial comoving radius of a shell (upper panel) and vs. $r$, the physical radius (lower panel) at $z=0$. Given the same linear amplitude today, the higher $w$ models have higher amplitude in the non-linear regime because of differences in the growth factors. If the density and infall velocity are higher the system gets more tightly bound, effectively causing a bunching and weakening of the $w$ dependence when plotted against the physical radius $r$.}
\label{densitytoday}
\end{figure*}

\capfigref{densitytoday} shows the final evolved density and infall velocity profiles at $z=0$ for three cases: $w=-1$ (red, dotted), $w=-1/2$ (green, dotdashed) and  $w=-3/2 $ (brown, long dashed). The runs here correspond to initial conditions with $\sigma_8=0.8$. The plots in the upper panel are 
drawn against the Lagrangian coordinate ($X$) and those in the bottom panel are drawn against the physical distance from the cluster centre ($r$). In the upper panel, the differences in evolution for different $w$ models become apparent in the non-linear regime and the curves merge in the linear regime. This is because the linear theory power is fixed to be the same in all models at $z=0$. Consequently, the amplitude of the initial profile at recombination is higher for higher $w$ models  because the growth factors are smaller and hence the final non-linear amplitude is more in these models. This is in qualitative agreement with results of \cite*{mcdonald_dependence_2006}, who examined  the effect of $w$ on the matter power spectrum. The sensitivity to $w$ is almost eliminated when the curves are plotted against the physical radius $r$. In the absence of perturbations, the physical radius is just the comoving radius multiplied by the expansion factor $r=a(t) X$. For collapsing perturbations the net physical radius is always less than this; the higher the density and infall velocity, the more tightly bound the system. This causes a bunching of the curves when plotted against $r$.  

\begin{figure*}
\includegraphics[width=16cm]{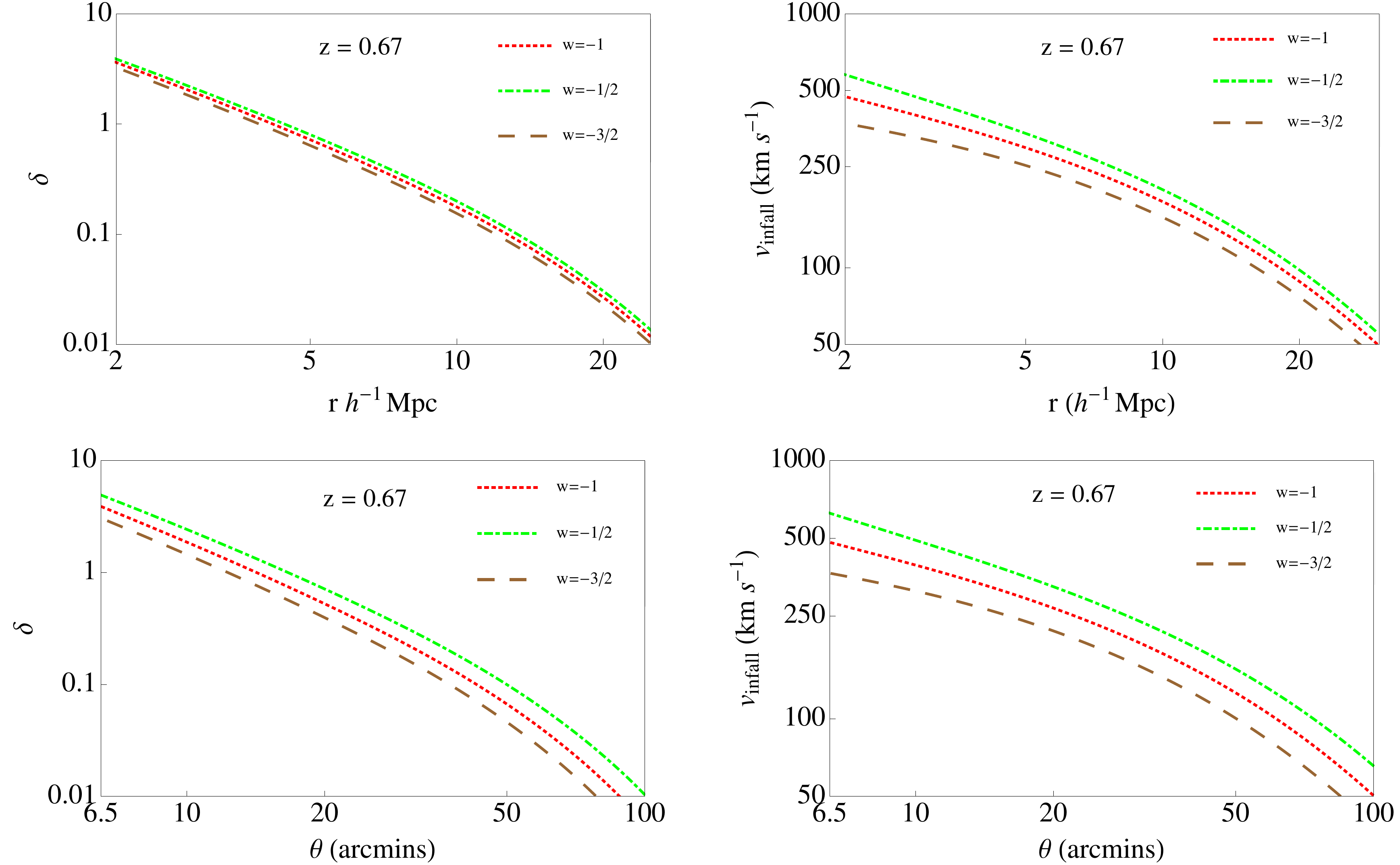}
\caption{Density and infall velocity profiles at redshift $z=0.67$. At higher redshifts, metric distance is replaced by the angular separation, which itself depends on the background cosmology. Plotting the profiles against the angular separation shows a increased deviation between models. However, isolating peculiar velocities at these redshifts is not possible with current observations.}
\label{densityearly}
\end{figure*}

The situation is not too different at higher redshifts. At higher redshifts the distance from the cluster centre $r$ (metric separation) is not the relevant observable and is replaced by the angular separation $\theta$. For a flat universe, the two are related as 
\beq
\theta = \frac{r }{a \chi(a)}.
\label{eq:thetavsr}
\eeq 
$\chi(a)$ is the comoving distance to the object defined as 
\beq
\chi(a) = c \int_a^1 \frac{da}{a^2 H(a)},  
\eeq
where $H(a)$ is given by \eqnref{eq:freid} and $c$ is the speed of light. 
\capfigref{densityearly} shows the density and infall velocity patterns at $z=0.67$ plotted against the metric separation (top panel) and the angular separation (bottom panel). At this redshift, for the fiducial $\Lambda$CDM model, \eqnref{eq:thetavsr} reduces to $\theta = 3.38 r$.
As was the case at $z=0$, the density and velocity profiles when plotted against the metric separation, are not too sensitive to the change in $w$. But the angular separation also depends on $w$. For the same metric separation, higher $w$ implies a larger angular distance. This broadens the curves and marginally improves the sensitivity to the equation of state parameter especially in the outer regions of the cluster. The redshift $z=0.67$ is chosen as a typical value in the range $z\sim 1$. In this range the Zel'dovich curves for different $w$ have the maximum spread (see \figref{fig:zeldovich}) and in the next section we show that they can be exploited to remove degeneracies due to parameters such as $\sigma_8$. 

Determination of individual peculiar velocities requires having redshift independent distance measures.
 Such measures, for e.g., the Tully-Fisher relation (\citealt{tully_new_1977}), are applicable only to low redshifts ($z\sim 0.1$). At these redshifts, they give errors on the order of $\sim$ 100-150 km ${\rm s}^{-1}$ (e.g., \citealt{davis_local_2011}) making it practically impossible to make any reliable estimates at high redshifts, where the error will be worse. Type Ia supernovae may be the best bet to determine individual peculiar velocities at high redshifts, but even these methods have been restricted to low redshifts (for e.g., $ z \sim 0.067$ in \citealt{turnbull_cosmic_2012}). Currently and in the near future only statistical information about high redshift peculiar velocity fields will be available via redshift space distortions .

\subsection{Parameter degeneracies and the phase space picture}
\begin{figure*}
\includegraphics[width=16.5cm]{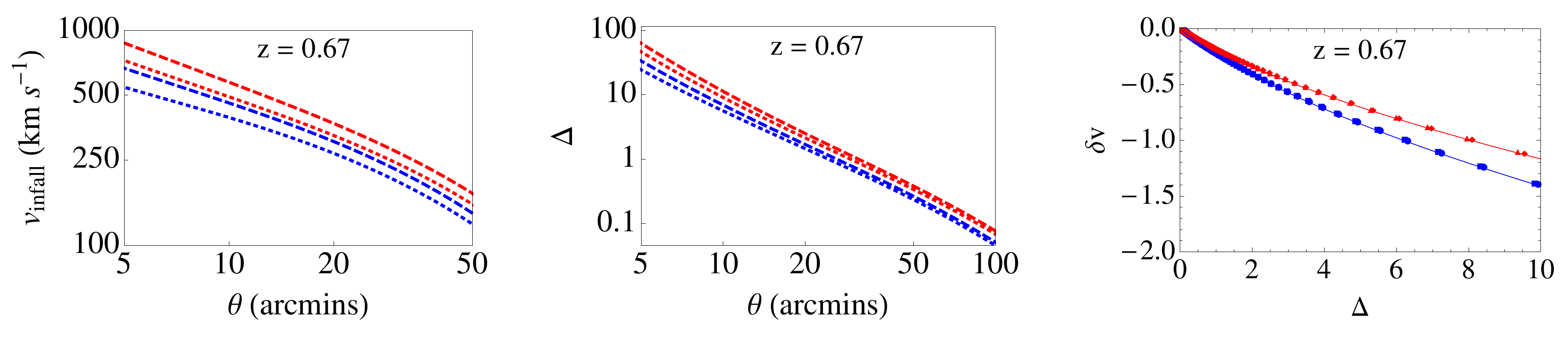}
\caption{Phase space plot as a tool to lift the degeneracy of parameters. Blue (red) lines are for $w=-1$ ($w=-1/2$) and dotted (dashed) lines are for $\sigma_8=0.8$ ($\sigma=0.9$). A higher value of $\sigma$ has the same effect as a higher value of $w$ and plotting only the infall velocity (first plot) or the spherically averaged overdensity (second plot) cannot constraint $w$. However, plotting the same information on the phase space portrait removes this degeneracy. The 
Zel'dovich curve, which is a relation between the spherically averaged density $\Delta$ and fractional Hubble parameter $\delta_v$, distinguishes cosmologies with different values of $w$. }
\label{deltadeltav}
\end{figure*}
The theoretically predicted profiles in \figref{densityearly} were obtained using fixed values of input parameters such as $\sigma_8$ (chosen to be 0.8).  Determination of 
$\sigma_8$ has its own observational uncertainties which introduces a degeneracy between $\sigma_8$ and $w$.
Thus, using only the density or only the velocity information cannot not constrain $w$ uniquely. We demonstrate that, at redshifts $z \sim 1$, this degeneracy can be removed by combining the density and velocity information by plotting it on the $\Delta-\delta_v$ phase space plot. 

\capfigref{deltadeltav} shows two sets of average density and velocity profiles and the phase space portrait at $z=0.67$. The lower blue set corresponds to $w=-1$ with $\sigma_8 =0.8$ (dotted) and $\sigma_8=0.9$ (dashed) and the upper red set is for $w=-1/2$ with the same $\sigma_8$ values. The velocity profiles for $w=-1, \sigma_8=0.9$ almost coincide with $w=-1/2, \sigma_8=0.8$. The same is true for the averaged density $\Delta$. But, when the velocity information is combined with the density information and plotted on the phase space portrait, the $w=-1$ and $w=-1/2$ points clearly separate and practically lie on their corresponding Zel'dovich curves. The relative deviation from the Zel'dovich curve was calculated for all four sets of points and the maximum over all the points was $1.6 \times 10^{-5}$. 

The attracting nature of the Zel'dovich curve ensures small uncertainties in the initial conditions do not affect the final state of the perturbations. Therefore, any degeneracy parameter that affects the initial conditions but not the evolution equations can be eliminated by plotting the density and velocity information on the phase space portrait. Theoretically, this will work best at redshifts near $z \sim 1$ where the spread in the Zel'dovich curves for $w$ is maximum. Unfortunately, as was discussed earlier, isolating peculiar velocities at such a high redshift is an extremely challenging task. In addition, degeneracy parameters such as galaxy bias, which arise from observational constraints, cannot be eliminated by this plot.

\section{Summary and conclusion}
\label{sec:conc}
We have obtained the non-linear density-velocity divergence relation by examining the dynamics of perturbations in the joint $\delta-\delta_v$ phase space. Although we were restricted to spherical top-hat models, unlike standard practice, we did not use exact solutions of the top-hat in our derivation of the result. Instead, at each epoch we obtained pairs of initial $(\delta_i, \delta_{v,i})$ which satisfied the condition `no perturbations at the big bang'. These pairs traced out a curve in the $\delta -\delta_v$ phase space which we refer to as the `Zel'dovich curve'.
Because of the non-autonomous nature of the $\delta-\delta_v$ evolution equations, this curve is a time dependent entity. We demonstrated that the curve acts like an `attractor' for the dynamics of the perturbations. Small amplitude perturbations, as they evolve through the phase space, asymptotically approach this curve and perturbations that start along the curve stay along it with a very high accuracy. Thus, the Zel'dovich curve gives the long term behaviour of density and velocity perturbations and is exact non-linear extension of the density-velocity curve for the spherical top-hat. 

We obtained a 3\% fit to the Zel'dovich curve in the range $-1\leq \delta \leq10$ by generalising existing formulae of B92 and BC08 to account for the inclusion of dark energy. From this we obtained a new parametrization for the linear growth index $\gamma = 0.56(-w)^{-0.08} -0.01 (-w)^{-1.18}$, which agrees with the well known result of \citet{linder_cosmic_2005} for $\Lambda$CDM models and deviates by at most $1\%$ for values of $w$ between $-3/2\leq w \leq -1/2$. As expected, the explicit dependence of the $\delta-\delta_v$ relation on $w$ is weak, both in the linear regime and in the non-linear regime. Nevertheless there is a implicit dependence on $w$ through the evolution of $\Omega_m$. 

As a practical application, we considered the evolution of density and velocity profiles of galaxies in the quasi-linear regime ($\delta \sim 5$), before effects of virialization become important, and investigated the dependence on $w$. We found that the deviations in cluster profiles due to a 50\% change in $w$ are about $\sim100$ km/s which are barely within the current observational errors in the local universe (\citealt{courtois_cosmic_2012}, \citealt{davis_local_2011}). We also demonstrated that the degeneracy in the density and velocity profiles arising due to parameters such as $\sigma_8$ can be broken if the same information is plotted on the $\delta-\delta_v$ plot. The attracting behaviour of the Zel'dovich curve makes it insensitive to small changes in the initial parameters and the points get classified according to $w$. However, this is works best only at redshifts $z\sim 1$, where the curves for different $w$ have the maximum spread. Given the observational difficulties in isolating peculiar velocities at these redshifts, this result probably is only of theoretical value at the moment. 

The main new feature of this work is the interpretation of the non-linear density-velocity divergence relation in the phase space picture and the generality of the method. In this paper we focussed only on the simplest possible phenomenological model for dark energy: constant equation of state, not necessarily $-1$. It would be interesting to consider more generalised models such as varying equation of state, coupled dark energy models or alternate models of gravity. Spherical collapse models have already been worked out for various dark energy or quintessence models (for e.g., \citealt{mota_spherical_2004}; \citealt{maor_spherical_2007}; \citealt{pace_spherical_2010}) and modified gravity scenarios (for e.g., \citealt*{dai_consequences_2008}; \citealt{schafer_spherical_2008}). 
The general method remains the same: obtain the correct equation for the evolution of the scale factor, impose the `equal age' condition to get constrains on the initial density and velocity parameters and examine the behaviour of the resulting curve in the phase space dynamics. However several potential caveats and questions may arise. For example in modified gravity scenarios the evolution of a spherical shell enclosing a fixed mass is not independent of the internal distribution of the mass or the environment; Birkoff's theorem does not hold true and spherical top-hats do not remain top-hats as the evolution proceeds (\citealt{dai_consequences_2008}; \citealt*{borisov_spherical_2012}). Clearly the $\delta-\delta_v$ relation will have to be generalised to treat such situations. It remains to be investigated whether there exist variables related to density and velocity that satisfy a unique relation in phase-space. Recent simulations (\citealt{schmidt_nonlinear_2009}; \citealt{lombriser_cluster_2012}) have shown that halo density profiles calculated in $f(R)$ models of modified gravity show an enhancement at a few virial radii when compared to those evaluated with `standard' $\Lambda$CDM. It would be interesting to investigate if the these features translate to a signature in some appropriate Zel'dovich relation. 

The Zel'dovich curves were obtained using a spherical top-hat model. In truly inhomogenous systems, the $\delta-\delta_v$ relation is no longer one to one and the joint probability distribution of the two variables provides a more complete description. Whether or not the Zel'dovich relation is obeyed even in a spherically averaged sense is not yet examined. Numerical simulations will be required to give more refined theoretical answers. From an observational point of view too, there are many sources of deviation from spherical symmetry; tidal interactions give rise to transverse forces, infall of smaller systems superimposes a random velocity on the radial infall (e.g., \citealt{diaferio_infall_1997}) etc. In addition to the difficulties associated with isolating peculiar velocities, these pose further constraints on the application of the Zel'dovich relation to real systems. Nevertheless, the Zel'dovich curve can provide a pivot point to analyse more complicated systems and we hope that its significance in the density-velocity phase space dynamics can be exploited to constrain cosmological parameters in the future.

\section*{Acknowledgements}
I would like to thank David F. Chernoff for helpful discussions as well as feedback on the paper and Martha Haynes, the ExtraGalactic Group at Cornell University, J.K. Bhattacharjee, Eanna Flanagan, Rachel Bean, Valeria Pettorino and Shubabrata Majumdar for useful conversations. In addition, I would like to thank Maciej Bilicki for suggestions and valuable comments on the manuscript.
%

%\clearpage
%\section{References}
%The style file is journal specific. 
\bibliographystyle{mn2e.bst}
\bibliography{mybibtex1,mybibtex2}

\appendix
\section{Evolution equations for the spherical profile}
\label{App:SCM}

This appendix gives the equations that govern the evolution of a spherically symmetric matter perturbation in a background cosmology comprising of dark matter and dark energy described by a constant equation of state $w$. Dark energy is not coupled to dark matter and is assumed to stay spatially uniform. The equations derived here hold for perturbations with arbitrary radial density and velocity profiles. The spherical top-hat of \S \ref{sec:dynamics} becomes a special case. 

The physical configuration of the compensated spherical perturbation consists of an inner sphere surrounded by a compensating spherical shell. The background extends beyond the outer edge of the compensating shell. The density in the compensating shell is adjusted so that the mean density of the total mass enclosed within the perturbation is the same as that of the background. Let $\rho_m$ (${\tilde \rho}_m$) denote the density of the background (perturbation). The origin is at the centre of the inner sphere and $r$ and $x$ denote the physical and comoving distance from the centre.

We start with the generalised Euler equation for the evolution of the velocity (\citealt{lima_newtonian_1997}; \citealt{abramo_structure_2007}; \citealt{pace_spherical_2010}) 
\bea
\label{eq:x}
\frac{\partial {\bf u}}{\partial t} + {\bf u} \cdot \nabla_x {\bf u} + 2 \frac{\dot a}{a}{\bf u}  &=& -a^{-2} \nabla_x \Phi, \\
\label{eq:phix}
\nabla_x^2  \Phi &=& 4 \pi G a^2 \delta \rho_m({\bf x}, t),
\eea
where ${\bf x}$ is the comoving coordinate, ${\bf u} = d {\bf x}/dt$, $\Phi$ is the peculiar gravitational potential and $\delta \rho_m$ is the matter density perturbation. There are two main frameworks to describe the evolution of a fluid: Eulerian and Lagrangian. 
In the Eulerian description, one generally solves \eqnref{eq:x} for ${\bf u}$ as a function of some fixed grid coordinates, whereas in  the Lagrangian description, the position is the main variable and is solved as a function of some initial coordinates and time. Since the main aim is to solve for the scale factor, the latter approach is more convenient. Combining the first two terms into a total derivative ($ \partial /\partial t + {\bf u} \cdot \nabla_x = d/dt$), using spherical symmetry and changing to physical radial coordinates $r=a x$, \eqnrefs{eq:x} and \eqnrefbare{eq:phix} become
\bea
\label{Appeq:r}
{\ddot r} - \frac{{\ddot a}}{a} r &=& - \nabla_r \Phi, \\
\nabla_r^2  \Phi &=& 4 \pi G \delta \rho_m,
\label{eq:phi}
\eea
where the dot is derivative w.r.t. time $t$ and the background scale factor evolves as 
\beq
\frac{\ddot a}{a} = -\frac{H_i^2}{2}  \left[\frac{\Omega_{m,i} a_i^3}{a^3} +(1+3w) \Omega_{\phi,i} \left(\frac{a_i}{a}\right)^{3(1+w)}\right].
\label{eq:appback}
\eeq

Define the Lagrangian coordinate of a shell as the initial comoving coordinate of the shell i.e., $X=r_i/a_i$. 
The physical radius of any shell at a later time can be written as 
\beq 
r(X,t) = b(X,t) X,  
\label{eq:eultolag}
\eeq
where $b(X,t)$ can be thought of as the scale factor of the shell at $X$. By definition $b(X,t_i) = a_i$. 
The perturbation is described by two quantities: initial overdensity parameter
\beq 
\delta(X,t_i) = \frac{{\tilde \rho}_m(X,t_i)}{\rho_{m,i}} -1
\label{eq:deltaXti}
\eeq
and the velocity perturbation parameter 
\beq 
\delta_v(X,t_i) = \frac{1}{H_i} \frac{{\dot r}(X,t_i)}{r(X,t_i)} -1, 
\eeq 
where ${\dot r}(X,t_i)$ and ${\tilde \rho}_m(X,t_i)$ are the initial total velocity and initial total density profiles of the perturbation and $\rho_{m,i}$ is the initial total density of the background. 
Conservation of mass implies that the matter density at any later time 
\beq 
{\tilde \rho}_{m}(X,t) = \frac{{\tilde \rho}_{m}(X,t_i)J(X,t_i)}{J(X,t) }, 
\label{eq:rhoXt}
\eeq
where $J$ is the Jacobian factor relating the Eulerian and Lagrangian volume elements,
\beq
r^2 dr = J(X,t) X^2 dX.
\label{eq:volelem}
\eeq 
Using \eqnref{eq:eultolag}, 
\beq
J(X, t)= b^3 \left(1+\frac{X}{b} \frac{db}{dX} \right).
\eeq
By construction, 
\beq J(X,t_i) = a_i^3.
\label{eq:Jatti}
\eeq 
The background density $\rho_{m} = \rho_{m, i} a_i^3/a^3$ and using \eqnrefs{eq:deltaXti}, \eqnrefbare{eq:rhoXt} and \eqnrefbare{eq:Jatti} the perturbed density evolves as 
\beq 
\delta \rho_{m}(X,t) =\frac{\rho_{m,i} a_i^3}{a^3}\left[ \frac{(1+\delta(X,t_i))a^3 }{J(X,t)}-1\right].
\label{eq:deltarho}
\eeq
From spherical symmetry, the force 
\beq
 \nabla_r \Phi =  \frac{d\Phi}{dr} = \frac{4 \pi G}{r^2} \int_0^r \delta \rho_m (r') r'^2 dr'.
\label{eq:force}
\eeq
Convert to Lagrangian coordinates by substituting \eqnrefs{eq:eultolag}, \eqnrefbare{eq:volelem} and \eqnrefbare{eq:deltarho} in  \eqnrefbare{eq:force} 
\beq
\nabla_r \Phi = \frac{4 \pi G\rho_{m,i} a_i^3 b X}{3 a^3} \left[ \frac{3 a^3}{b^3 X^3}\int_0^X (1+\delta(X',t_i)) X'^2 dX'.  -1 \right].
\label{phiX}
\eeq
Define
\beq
\Delta(X,t_i) = \frac{3}{X^3} \int_0^X (\delta(X',t_i)) X'^2 dX'.
\eeq
$\Delta(X,t_i)$ is the average fractional density of mass within radius $X$. 
Substituting \eqnrefs{eq:eultolag} and \eqnrefbare{phiX} in \eqnref{Appeq:r}, setting $4 \pi G \rho_{m,i} /3 = 1/2 H_i^2 \Omega_{m,i}$ and using \eqnref{eq:appback} for the evolution of the background gives %
\beq
\frac{{\ddot b}}{b} = -\frac{H_i^2}{2} \left[\frac{\Omega_{m,i} a_i^3 (1+\Delta(X,t_i))}{b^3}  + (1+3w) \Omega_{\phi,i} \left(\frac{ a_i}{a} \right)^{3(1+w)}\right].
\label{eq:generalb}
\eeq
The initial conditions are $b(X,t_i) = a_i$ and ${\dot b}(X,t_i) = {\dot a}_i(1+ \delta_v(X,t_i))$. 

If the initial density is uniform, $\Delta(X, t_i) = \delta_i$ and  $\delta_v(X,t_i) = \delta_{v,i}$ and the $X$ dependence drops out of the equation for evolution of $b(X,t)$. An initially uniform spherical perturbation continues to stay uniform and can be described by $r(X,t) = b(t)X$. 
\beq
\frac{{\ddot b}}{b} = -\frac{H_i^2}{2} \left[\frac{\Omega_{m,i} a_i^3 (1+\delta_i)}{b^3}  + (1+3w) \Omega_{\phi,i} \left(\frac{ a_i}{a} \right)^{3(1+w)}\right],
\label{eq:speciallb}
\eeq
with initial conditions $b(t_i) = a_i$ and ${\dot b}(t_i) = {\dot a}_i(1+ \delta_{v,i})$. This equation is similar to many others stated more directly in the literature (see for e.g., \citealt{percival_cosmological_2005}). Here we use the route advocated in recent papers (\citealt{pace_spherical_2010}; \citealt{wintergerst_clarifying_2010}) so that the set up can be generalised to cases where dark energy may have more complicated behaviour.

In the above derivation we have assumed that the Lagrangian system is a good coordinate system to describe the evolution. This is true as long as no shell crossing occurs within the system, which ensures that the mapping between the physical coordinates $r$ and the Lagrangian coordinates $X$ is unique. For the top-hat case, the argument that supports this assumption is presented in Appendix A of NC. Here we extend it to the spherical infall model. Let $r_{edge,i}$ be the initial physical distance of the edge of the cluster from the center. The velocity of the edge is ${\dot r}_{edge,i} = (1+\delta_v(X_{edge}, t_i)) H_i r_{edge,i}$. Given the perturbation parameters $\delta(X_{edge},t_i)$ and $\delta_v(X_{edge}, t_i)$, one can always choose $r_{edge,i}$ (and hence ${\dot r}_{edge,i}$) arbitrarily small so that the time for the edge to reach any physical distance is arbitrarily large. 
The initial density $\delta(X,t_i)$ is maximum at the center and decreases monotonically as $X$ increases (see Appendix \ref{app:initden} for the 
initial set-up). The velocity perturbation $\delta_v(X,t_i)$ is proportional to $\delta(X,t_i)$, but with a negative sign and hence $(1+\delta_v)$ increases as one moves from the center to the edge. Thus, the net gravitational binding is tighter near the cluster center than the edge.  This ensures that a shell initially closer to the center does not cross any shell at a greater initial radius and collapses into the center earlier. The choice of the edge radius fixes the net mass inside. The additional mass needed to satisfy mass conservation is put in a compensating shell between the cluster edge and the inner edge of the background and set on a critical trajectory outward so that it moves along with the background. 

\section{Time dependence of the Zel'dovich condition for dark energy}
\label{App:zellcdm}

The evolution of the background and perturbation scale factors for the top-hat perturbation is given by 
\bea
\frac{\ddot a}{a} &=& -\frac{H^2_i}{2} \left(\frac{\Omega_{m,i} a_i^3}{a^3} + (1+ 3 w) \Omega_{\phi,i} \left(\frac{a_i}{a}\right)^{3(1+w)}\right) 
\label{backscalefac2}\\
\frac{\ddot b}{b} &=& -\frac{H^2_i}{2} \left(\frac{\Omega_{m,i} a_i^3 (1+ \delta_i)}{b^3} + (1+ 3 w) \Omega_{\phi,i} \left(\frac{a_i}{a}\right)^{3(1+w)}\right), 
\label{pertscalefac2}
\eea
where $\Omega_{m,i}, \Omega_{\phi,i} $ are the Hubble parameter and density parameters at the initial time $t_i$ and are related to their values today ($a=0$) through 
\bea
\Omega_{m,i} &= & \frac{\Omega_{m,0} a_i^{-3}}{\Omega_{m,0} a_i^{-3} + \Omega_{\phi,0} a_i^{-3(1+w)}},  \\
\Omega_{\phi,i} &=&   \frac{\Omega_{\phi,0} a_i^{-3(1+w)}}{\Omega_{m,0} a_i^{-3} + \Omega_{\phi,0} a_i^{-3(1+w)}}.  
\label{omegachange2}
\eea
The initial conditions are $a(t_i) = a_i$, ${\dot a}(t_i) =H_ia_i$, $b(t_i) = a_i$ and ${\dot b}(t_i) = H_ia_i(1+ \delta_{v,i})$. 
Substitute $x= a/a_i$,$y = b/a_i$, $\tau= t_i H_i$ in \eqnrefs{backscalefac2} and \eqnrefbare{pertscalefac2}. 
The two equations then read 
\bea
\label{xeq}
\frac{1}{x} \frac{ d^2 x}{d\tau^2} &= &-\frac{1}{2} \left(\frac{\Omega_{m,i}}{x^3} + (1+ 3 w) \Omega_{\phi,i} \left(\frac{1}{x}\right)^{3(1+w)}\right),\\
\label{yeq}
\frac{1}{y}\frac{d^2 y}{d\tau^2} &= &-\frac{1}{2} \left(\frac{\Omega_{m,i} (1+ \delta_i)}{y^3} + (1+ 3 w) \Omega_{\phi,i} \left(\frac{1}{x}\right)^{3(1+w)}\right),
\eea
with initial conditions $x(\tau_i) = 1$, $\frac{dx}{d\tau}(\tau_i) = 1$, $y(\tau_i) = 1$, $\frac{d y}{d\tau}(\tau_i) = (1+ \delta_{v,i})$. Note that the equal age condition is unchanged by this scaling since the time is scaled by the same constant both for the background and perturbation. It is also clear that the equal age condition will only involve the parameters $\Omega_{m,i},\Omega_{\phi,i}$ and $w$. For a flat EdS universe $\Omega_{\phi,i}=0$ and $\Omega_{m,i} =1$ at all times. 
 \capeqnrefs{xeq} and \eqnrefbare{yeq} can be integrated to give
\beq 
\left(\frac{dx}{d\tau}\right)^2 = \frac{1}{x}; \mbox{ 		} \left(\frac{dy}{d\tau}\right)^2 = \frac{1+\delta_i}{y} + (1+ \delta_{v,i})^2 -(1+\delta_i).
\eeq
The age of the background (perturbation) is the time elapsed for $x(y)$ to grow from 0 to 1. Imposing the `equal age' condition, gives an implicit relation between $\delta_i$ and $\delta_{v,i}$
\beq
\int_{y=0}^{y=1}\frac{dy}{\left[(1+\delta_i) y^{-1} +  (1+ \delta_{v,i})^2 - (1+\delta_i) \right]^ {1/2}} = \frac{2}{3}.
\eeq

In the presence of a dark energy term analytic calculations are not possible. For such cases, given a $\delta_i$ and $\delta_{v,i}$ at $\tau=\tau_i$, the age of the perturbation can be computed by integrating \eqnref{yeq} back in time until $y$ reaches zero. Fixing $\delta_i$ one performs a search in the $\delta_{v,i}$ parameter space until the age of the background and perturbation are the same. However for models with $w<-1$, an additional complication arises because \eqnref{yeq} has a singularity when $x=0$. This condition is independent of $\delta_{v,i}$ and can complicate the search. So, instead of backward integration, we set $y\approx0$ at the big bang time (epoch when $x=0$) and integrate forward in time until $\tau_i$. The time elapsed gives the age of the perturbation. A search is performed to find the velocity at the bang time for which the ages of the background and perturbation are the same. The initial velocity at $\tau_i$ can be read off from the solution: $\delta_{v,i} = {\dot y}(\tau_i) -1$. 
\section{Setting the initial density profile of cluster progenitors}
\label{app:initden}
The initial density and velocity profiles are set, as per the prescription in LL91, with some minor modifications. This paper uses results from the BBKS (\citealt{bardeen_statistics_1986}) analysis. The basic premise is that matter collapses around the peaks in the primordial density field, filtered on some appropriate length scale, forming the progenitors of bound cosmic structures. 

Start with the CDM power spectrum $P(k)  = P_0 k^{n_s} T^2(k)$, where the transfer function has the BBKS form (\citealt{bardeen_statistics_1986})
\bea 
T(q) &=& \frac{ln(1+2.34q)}{2.34 q} \left[1+3.89q + (16.1q)^2 + (5.46q)^3 + (6.71 q)^4\right]^{-1/4}, \\
q & =& k/(\Omega_m h^2 {\rm Mpc}^{-1}). 
\eea
$k$ is measured in $h {\rm Mpc}^{-1}$. The spectral index $n_s$ was chosen, in accordance with LL91 to have the Harrison-Zel'dovich value $n_s=1$, which is slightly higher than the current WMAP `standard' value from seven year observations $n_s = 0.967$ (\citealt{komatsu_seven-year_2011}). The normalisation $P_0$ is set by demanding that the r.m.s. fluctuation on 8 $h^{-1}$ Mpc ($\sigma_8$) defined as 
\beq 
\sigma_8^2 = \frac{1}{2 \pi^2} \int_0^{\infty} k^2 P(k) \left[\frac{3(\sin ka - ka \cos ka)}{(ka)^3}\right]^2 dk, 
\eeq
where $a= 8 h^{-1} {\rm Mpc}$, matches the observed value. 
We ran separate cases with two values $\sigma_8=0.8$ and $\sigma_8=0.9$. The normalised power spectrum is filtered with a Gaussian filter of width $3 h^{-1} {\rm Mpc}$ 
\beq 
P_f(k) = P(k) W^2(R_f,k),
\eeq
where 
\beq 
W(R_f,k) = e^{-\frac{1}{2}(R_fk)^2}.
\eeq
The height of any peak in the density distribution is usually denoted as $\nu \sigma_0$ where $\sigma_0$ is the r.m.s. of the filtered density field. The ensemble average of the radial distribution of the density field around a peak of relative height $\nu$ is given by 
\beq
\delta(r, \nu) = \frac{1}{2 \pi^2 \sigma_0} \int_0^\infty \frac{\sin kr}{kr} \left[\frac{\nu -\gamma^2 \nu -\gamma \theta}{1-\gamma^2} + \frac{\theta R^2_*}{3\gamma (1-\gamma^2)} k^2 \right] dk.   
\label{deltainit}
\eeq
Here $\gamma$ and $R_*$ are spectral parameters and $\theta$ is a function of $\nu$ and $\gamma$ (see LL91).

The fractional peak height $\nu$ is a random variable with a comoving differential number density $N_{pk}(\nu)$ and ideally one must evolve many initial profiles with different values of $\nu$ drawn from this distribution. Instead, we follow LL91 and substitute for $\nu$ and $\theta$ in \eqnref{deltainit} the average values $\langle \nu \rangle$ and $\langle \theta \rangle$
\beq 
\langle \nu \rangle = \frac{\int_{\nu_t}^\infty \nu {\mathcal N}_{pk}(\nu) d\nu }{\int_{\nu_t}^\infty {\mathcal N}_{pk}(\nu) d\nu }; \mbox{	     } \langle \theta \rangle = \frac{\int_{\nu_t}^\infty \theta(\nu, \gamma) {\mathcal N}_{pk}(\nu) d\nu }{\int_{\nu_t}^\infty {\mathcal N}_{pk}(\nu) d\nu },
\eeq
where the threshold $\nu_t$ is the peak height above which clusters can form. This is set by requiring that the theoretical number density of galaxy clusters above the threshold equals the observed number density $n_{pk,obs}$.
\beq 
n_{pk}(\nu_t) = \int_{\nu_t}^\infty {\mathcal N}_{pk}(\nu) d\nu. 
\eeq
 We use $n_{pk, obs} \approx 10^{-5} h^{3}{\rm Mpc}^{-3}$ for clusters with mass greater than $10^{14} M_{\odot}$ (\citealt*{allen_cosmological_2011}). 

The function $\delta(r, \langle \nu \rangle)$ set up by this method corresponds to the linear profile at $a=1$. This is transformed into the profile at any other redshift by multiplying by the appropriate growth factor. 
\beq 
\delta(r,\langle \nu \rangle)_{a=a_i} = \delta(r, \langle\nu\rangle)_{a=1} \frac{D(a=a_i)}{D(a=1)}, 
\label{deltar}
\eeq
where the growth factor $D(a)$ is (\citealt{heath_growth_1977}; \citealt{dodelson})
\beq
D(a) = \frac{5}{2} \Omega_{m,0} \frac{H(a)}{H_0}\int_0^a\frac{da'}{(a'H(a')/H_0)^3}.
\eeq
This expression for the growth factor has been derived for pressureless matter universes, but is also valid for dark energy cosmologies described by constant $w$. \S \ref{subsec:eq} sets up the initial profile in terms of Lagrangian coordinates: $\delta(X,t_i)  = \delta(r,\langle \nu \rangle)_{a=a_i} $. The initial profiles are chosen so that at $z=0$ the linear theory profile is same in all models. Since the growth factor is different for different $w$, the initial profiles have different amplitudes; higher $w$ values have slower growth i.e., $D(a=1)/D(a=a_i)$ is smaller and hence a larger initial amplitude. 
The initial average density perturbation inside a radius $X$ is 
\beq
\Delta (X,t_i ) = \frac{3}{X^3} \int_0^X X'^2 \delta(X', t_i) dX'.
\eeq
The velocity perturbation is chosen by imposing Zel'dovich initial conditions
\beq 
\delta_v(X,t_i)  = -\frac{1}{3} \Omega_{m,i}^{0.55} \Delta (X,t_i ). 
\eeq
At recombination, $\Omega_{m,i} \approx 1$ for all values of $w$. 
Note that the definition of $\delta_v$ slightly differs from a similar velocity perturbation parameter $\alpha$ defined in LL91.

\end{document}